\documentclass[fleqn,usenatbib]{mnras}

\usepackage{newtxtext,newtxmath}

\usepackage[T1]{fontenc}
\usepackage{ae,aecompl}

\usepackage{xcolor}

\usepackage{graphicx}	
\usepackage{amsmath}	
\usepackage{amssymb}	

\title[Dust dynamics in protoplanetary discs]{Inner dusty regions of protoplanetary discs - II. Dust dynamics driven by radiation pressure and disc winds}

\author[D. Vinkovi\'{c} and M. \v{C}emelji\'{c}]{
Dejan Vinkovi\'{c},$^{1}$\thanks{E-mail: dejan@iszd.hr (DV), miki@camk.edu.pl (M\v{C})}
and Miljenko \v{C}emelji\'{c},$^{2,3,4}$
\\
$^1$Science and Society Synergy Institute, Bana Josipa Jela\v{c}i\'{c}a 22,HR-40000, \v{C}akovec, Croatia\\
$^2$Nicolaus Copernicus Astronomical Center, Polish Academy of Sciences, Bartycka 18, 00-716 Warsaw, Poland\\
$^3$Academia Sinica, Institute of Astronomy and Astrophysics, P.O. Box 23-141, Taipei 106, Taiwan\\
$^4$Shanghai Astronomical Observatory, Chinese Academy of Sciences, 80 Nandan Road, Shanghai 200030, China
}

\date{Accepted XXX. Received YYY; in original form ZZZ}

\pubyear{2020}

\begin{document}
\label{firstpage}
\pagerange{\pageref{firstpage}--\pageref{lastpage}}
\maketitle

\begin{abstract}
We explore dust flow in the hottest parts of protoplanetary discs using the forces of gravity, gas drag and radiation pressure. Our main focus is on the optically thin regions of dusty disc, where the dust is exposed to the most extreme heating conditions and dynamical perturbations: the surface of optically thick disc and the inner dust sublimation zone. We utilise results from two numerically strenuous fields of research. The first is the quasi-stationary solutions on gas velocity and density distributions from mangetohydrodynamical (MHD) simulations of accretion discs. This is critical for implementing a more realistic gas drag impact on dust movements. The second is the optical depth structure from a high-resolution dust radiation transfer. This step is critical for a better understanding of dust distribution within the disc. 
We describe a numerical method that incorporates these solutions into the dust dynamics equations. We use this to integrate dust trajectories under different disc wind models and show how grains end up trapped in flows that range from simple accretion onto the star to outflows into outer disc regions. We demonstrate how the radiation pressure force plays one of the key roles in this process and cannot be ignored. It erodes the dusty disc surface, reduces its height, resists dust accretion onto the star, and helps the disc wind in pushing grains outwards. The changes in grain size and porosity significantly affect the results, with smaller and porous grains being influenced more strongly by the disc wind and radiation pressure.

\end{abstract}




\section{Introduction}

Understanding dust evolution in protoplanetary discs, as one of the pillars of planet formation, has attracted considerable research efforts \citep{haex}.
The inner disc regions, where the dust grains reach their sublimation point, is of a special interest due to its relevance to the physics of terrestrial planet formation \citep{Wolf} and to the study of dust mixing and thermal processing of grains \citep{Brownlee}. Unfortunately, the inner discs are difficult to resolve by direct imaging and we have to rely on interferometric, spectral and photometric observations that are often difficult to interpret \citep{Dullemond}.

The main obstacle is the plethora of complex physical phenomena happening simultaneously in the inner disc, which results in a highly dynamic environment. This is revealed in the spectral and photometric variability all along the spectrum, indicating various activities close to the stellar surface due to magnetic fields that extends into the disc, stellar flares, irregularities in accretion, turbulence in the disc, circumstellar obscuration due to transiting dust clouds, jets and outflows, physical and chemical evolution of dust, etc. \citep[e.g.][]{Brownlee,Cody,Rice,Bally,Kospal}.

Not surprisingly, theoretical modelling of such inner dusty discs is difficult and one of the shortcomings is inadequate handling of dust dynamics \citep{Haworth}. This means that dust is typically described as coupled with the gas, where the gas dynamics dictates the dust distribution. We know that this cannot be entirely correct, especially in the optically thin disc surface, where the gas density drops to the levels of dynamical decoupling of gas and dust. On the other hand, it has become evident that simple models of duty disc geometries cannot explain anomalously high levels of near infrared flux \citep{Vinkovic14} or transport of grains from inner to outer parts of the disc \citep{Brownlee}.

This led to attempts to incorporate mangetohydrodynamical (MHD) disc winds into the dust distribution modelling \citep{BansWind,Giacal,Miyake,Liffman20}. The wind drives gas outflow that could drag dust particles out of the disc surface into the optically thin, low density zone above the disc. This is an improvement in the inner disc modelling, but the implemented disc wind models have relied on analytical descriptions or simplified seady state solutions. In our work we make a step further and incorporate results of numerical MHD models into the equations for dust dynamics. 

The solutions in vertical and radial directions were derived separately in the first models of accretion disc \citep{PrB68,SS73} and many subsequent numerical and analytical works. Purely hydrodynamic analytical solution for a viscous thin accretion disc in three dimensions (3D) was given in \cite{Kita95} \citep[also][]{KK00}, following the method of asymptotic approximation first employed in \cite{Urpin84}. In this method, the physical quantities in the disc are expanded in Taylor series and solved separately in terms of the different orders of a small parameter $epsilon$, the disc thickness. The magnetic extension of such solution, given in \cite{CKP19}, produced only general analytical conditions which need to be satisfied for validity of the solution. The closure of the magnetic equations could be done only with input from numerical simulations, reaching a quasi-stationary state.

Here we use some of the quasi-stationary states from numerical solutions in \cite{Cem19} as a background to which we add the dust grains of different radii. We study the distribution of grains in background flows obtained with different parameters in the simulations. The numerical solutions which we used rely on alpha-viscosity prescription. More physically motivated would be the use of solutions with viscosity produced by the magneto-rotational instability (MRI) like in \cite{Flock13}, or solutions which would include some of the microphysics in the disc as in \cite{Baietal16,Bai16,Bai17} or thermochemistry \citep{Wangetal19}. We still use the alpha-viscosity because of the simpler relation to the available analytical solutions and to a great body of previous work related to the alpha-viscous disc. Our results can serve as a starting point for future research, which will inevitably include the disc microphysics and radiative transfer. Another advantage of using solutions from \cite{Cem19} is that it provides a parameter study, so that the results in the different parts of the parameter space could be compared.

We also put focus on the role of radiation pressure in the dust dynamics. This force is unavoidable in the case of small grains, but it has been ignored in previous similar studies. \cite{Vinkovic09} showed that a significant contribution to radiation pressure force can come from the hot dust close to the star, in this work we will focus only on the stellar component to this force, while the diffuse part will be addressed in the next paper in the series. These calculations require estimates on the dust optical depth, for which we use a high-resolution multigrain radiative transfer from our previous work in this series \citep{Vinkovic12} and results from the disc models by \citet{Flock16}.

In Section \ref{Sec:MHD} we describe the MHD models used in our dust dynamics equations, while in Section \ref{sec:opticaldetph} we describe the method of adding the radiative transfer results into our equations. 
Calculation of the radiation pressure force is explained in Section \ref{sec:RadPress}. These steps lead to equations that describe dust grain motion in Section \ref{sec:equations}. Section  \ref{sec:results} presents the numerical algorithm and results of solving the equations for different MHD models and grain types. 
In Section \ref{sec:discussion} we discuss how the obtained results complement other studies of inner discs. Our results are summarised in Section \ref{sec:conclusion}. 

\section{MHD models of T Tau winds}
\label{Sec:MHD}

The first global 3D radiation MHD simulations of an accretion disc with magneto-rotational instability were performed in \cite{Flock13}. In this proof-of-concept paper they captured distribution and abundance of the dust particles in the simulations. The use of ideal-MHD approach and problems with the grain sizes $\le 1\mu$m pointed out the novelty of such simulations. To perform a parameter study with smaller grain particles and use non-ideal MHD approach, we rely on analytical solution by \cite{KK00} for a 3D hydrodynamic thin accretion disc with Shakura-Sunyaev $\alpha$-prescription for the viscosity. We use such solution as an initial condition to which a stellar dipole magnetic field is added. With the addition of anomalous Ohmic resistivity in the disc, we obtained quasi-stationary solutions for a magnetized thin accretion disc.

Numerical simulations of star-disc magnetospheric interaction used in our study are described in detail in \cite{Cem19}. They are repeating, with minor amendments, the setup with initial and boundary conditions from \cite{ZF09}. The results are similar to \cite{R09} and references therein. Here we give only a brief description of the simulation setup. 

The disc is set with the initial conditions from \cite{KK00}. The initially non-rotating hydrostatic corona and poloidal magnetic field are added to such a disc, with the use of constrained transport for ensuring $\nabla\cdot{\mathbf B}=0$, and the split-field method for the evolution of the magnetic field. In this method, only changes from the initial stellar magnetic field are computed in a simulation \citep{Tan94,Pow99}. Parameterisation of the viscosity and resistivity is defined by \citet{SS73} $\alpha$-prescription as $\alpha_{\rm v} c^2/\Omega_{\rm K}$, where $\alpha_{\rm v}$ is a free parameter of viscosity, smaller than unity, and $\Omega_{\mathrm K}=\sqrt{GM_\star/r^3}$ is the Keplerian angular velocity at the cylindrical radius $r$. We also use $\alpha_m$ as a free parameter defining the Ohmic resistivity. Together with $\alpha_v$, it defines a magnetic Prandtl number $P_m={2\alpha_v}/{3\alpha_m}$. In our simulation both $\alpha_v$ and $\alpha_m$ follow the same geometric profile of the thin disc, proportional to the ratio of the sound speed and Keplerian angular velocity $C_s^2/\Omega_K$.
We assume that all the disc heating due to viscous and Ohmic dissipation is radiated away from the disc. Then the diffusive terms remain only in the equation of gas motion and the induction equation -- they are not present in the energy equation.

The code works in normalised units, so that the results can be rescaled to different stellar objects, as tabled in \cite{Cem19}. Here we refer to the case of Classical T Tauri Stars (CTTS) with the stellar mass $M_*=0.5M_\odot$ and radius $R_*=2R_\odot$.  However, the MHD simulations do not involve the stellar temperature and luminosity, which are the key parameters in the radiation pressure analysis. We extract this information by matching our stellar properties with the statistics of observed luminosities of T Tau stars as follows. 

The total luminosity is a sum of stellar $L_*$ and accretion luminosity $L_{acc}$:
\begin{equation}\label{eq:Ltot}
L_{tot}=L_* + L_{acc} = L_* + 0.318\chi \left(\frac{M_*}{M_\odot}\right)\left(\frac{R_\odot}{R_*}\right)\left(\frac{\dot{M}_{acc}}{10^{-8}M_\odot/yr}\right)
\end{equation}
where we used the expression for accretion luminosity from \citet{accretion} and the accretion rate $\dot{M}_{acc}$. With their suggested value $\chi=0.8$ and our stellar properties, we get
\begin{equation}\label{eq:Lacc}
    L_{acc}=6.4\times 10^{-3} \left(\frac{\dot{M}_{acc}}{10^{-9}M_\odot/yr}\right) L_\odot
\end{equation}
Their accretion simulations indicate that the "effective temperature" of the shock on the stellar surface is $T_{acc}\sim 8 000K$. The stellar luminosity can be estimated from properties of a sample of T Tau stars. We used data from \citet{Herczeg} and extrapolated $L_*=0.66 L_\odot$ using our $M_*$ and $R_*$. This implies the stellar temperature $T_*\sim 3700$K, which are all properties of a typical classical T Tau star of the spectral type K7-M0. 
If we approximate the spectra with the black body emissions then we can easily postulate a normalised synthetic stellar spectrum as
\begin{equation}\label{eq:fstar}
f^{star}_\lambda=\gamma b_\lambda(T_*) + (1-\gamma)b_\lambda(T_{acc})
\end{equation}
where $b_\lambda$ is the normalized Planck function and $\gamma=L_*/L_{tot}$. 

The MHD simulations are performed on a polar grid with the radial points distributed logarithmically between $r=1 R_*$ and $30 R_*$, while the angular grid is uniform. The simulations work in scaled units described by \citet{ZF09} - the gas density, time and velocity scales are, respectively
\begin{equation}
\begin{split}
    \rho_{d0}&=6\times 10^{-8} kg/m^3\\
    t_0&=6394\left(\frac{M_*}{0.5M_\odot}\right)^{-1/2}\left(\frac{R_*}{2R\odot}\right)^{3/2} \,\,{seconds}\\
V_{K*}&=2.18\times 10^5 \left(\frac{M_*}{0.5M_\odot}\right)^{1/2}\left(\frac{R_*}{2R\odot}\right)^{-1/2}\,\, m/s
\end{split}
\end{equation}

We explored various disc wind models that differ by several parameters: the viscosity parameter $\alpha_{\rm v}$, the disc resistivity $\alpha_{\rm m}$, the stellar dipole magnetic field at the stellar surface on equator $B_*$ and the stellar rotation period $P_*$. 
The wind velocity and density structure close to the disc can differ dramatically between models. The temporal evolution of the wind can also be very different - from very dynamic to almost steady. 

In order to illustrate how these differences in wind solutions affect the dust particles close to the inner disc edge, we selected three wind models as templates. They enable us to conceptualise different trajectories that dust particles can experience thanks to the gas drag and radiation pressure force. We chose snapshots from simulations exhibited in \citet{Cem19}, which reached quasi-stationary states after few tens of stellar rotations. Variations of the dynamical quantities from the average values in the interval of about 10 stellar rotations from which each snapshot is taken \citep[as shown in Fig-1 in][]{Cem19} were very small, well below 10\% of the total value. We took snapshots at the middle of each such interval, typically after about 70 stellar rotations.

In Table \ref{tbl:parameters} we show values of the parameters for models under consideration (named as model-A, B and H). The table also shows the mass accretion rates calculated from the MHD wind solution and the accretion luminosity derived from equation (\ref{eq:Lacc}). 
Even though accretion luminosities in our examples are small and could be essentially ignored, we keep them for the completeness of our approach as other authors might have situations with a significant accretion contribution and can still follow our methodology. 

The models A and B are chosen to represent runs with different magnetic field strengths and stellar rotation rates while keeping the viscous coefficient $\alpha_v=1$ constant. In these two cases, accretion flow in the disc is directed towards the star all across the disc height, throughout the whole disc. In the third case, model-H with $\alpha_v = 0.4$, there is a backflow in the disc, which introduces an additional component of the disc flow in the model. Such solutions are obtained in the purely hydrodynamical analytical solutions with $\alpha_v <= 0.68$, and appear also in the magnetic simulations \citep{RMishraetal20}. They would also include the cases with $\alpha_v <= 0.1$, which is considered most realistic from the numerical simulations incorporating magneto-rotational instability.

\begin{table}
 \caption{Parameters of different disc wind models used in this study.}
 \label{tbl:parameters}
 \begin{tabular}{ccccccc}
  \hline
  model & $\alpha_V$ & $\alpha_m$ & $B_*$ & $P_*$ & $\dot{M}_{acc}$ & $L_{acc}$\\
        &            &            &  kG & day& 10$^{-9}M_\odot/y$ & $L_\odot$ \\
  \hline
  model-A & 1 & 1 & 0.50 & 4.65 & 7.4 & 0.05\\
  model-B & 1 & 1 & 0.75 & 3.10 & 1.2 & 0.008\\
  model-H & 0.4 & 1 & 0.50 & 4.65 & 3.4 & 0.02\\
  \hline
 \end{tabular}
\end{table}

\begin{figure}
 \includegraphics[width=\columnwidth]{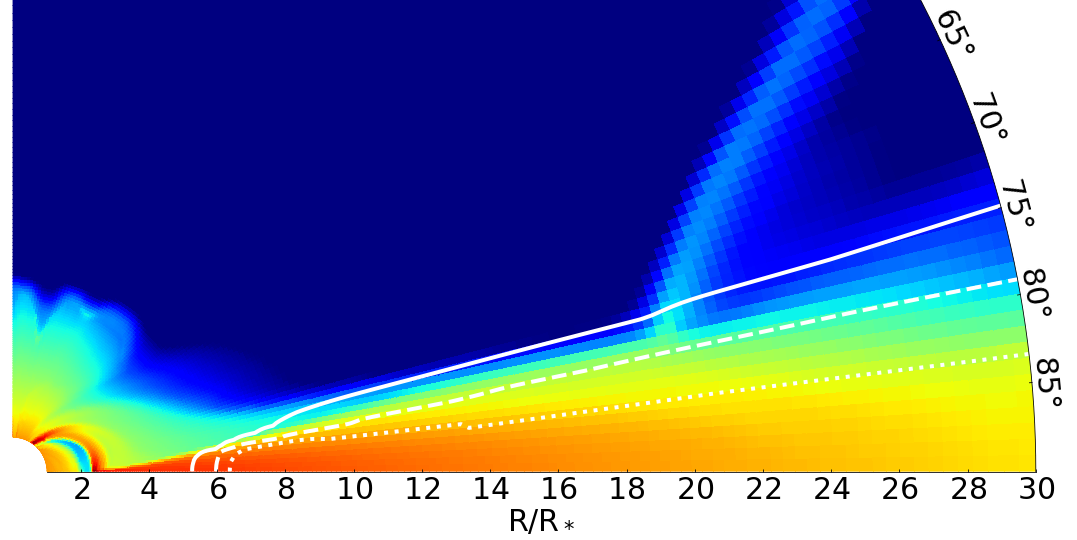}
 \includegraphics[width=\columnwidth]{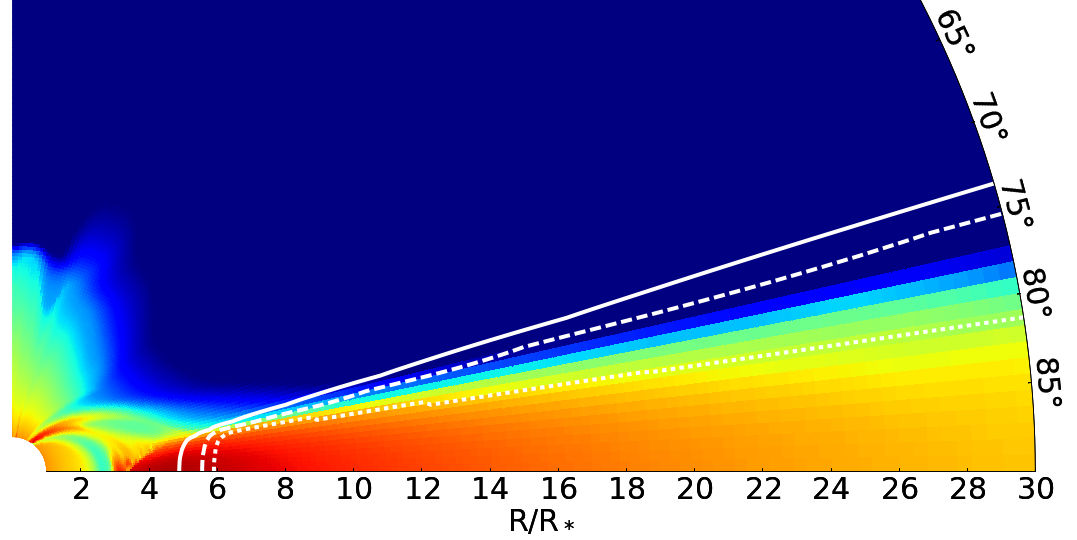}
 \includegraphics[width=\columnwidth]{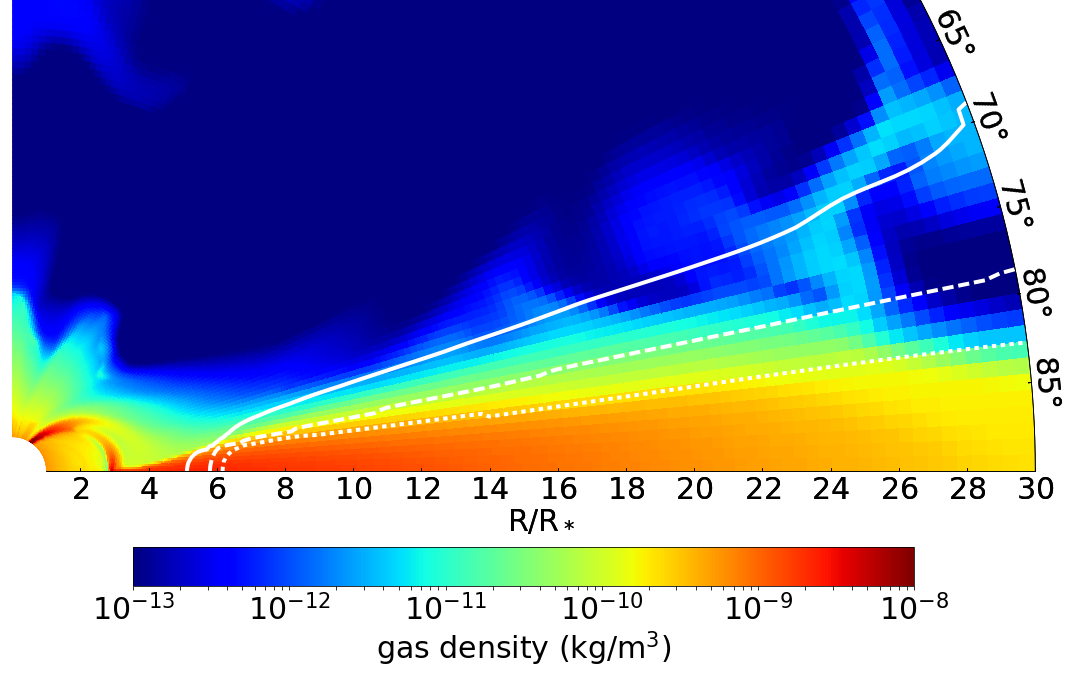}
 \caption{Gas density of disc wind models: the {\it upper panel} shows the result for model-A, the {\it middle} for model-B and the {\it lower} for model-H. The lines are contours of the optical depth structure (see equation \ref{eq:f}), where the {\it solid line} is $|\bmath{\varepsilon}|=0.99$, the {\it dashed line} is $|\bmath{\varepsilon}|=0.7$ and the {\it dotted line} is $|\bmath{\varepsilon}|=0.1$ (essentially the surface of optically thick disc). }
 \label{fig:GasDensity}
\end{figure}

\begin{figure}
 \includegraphics[width=\columnwidth]{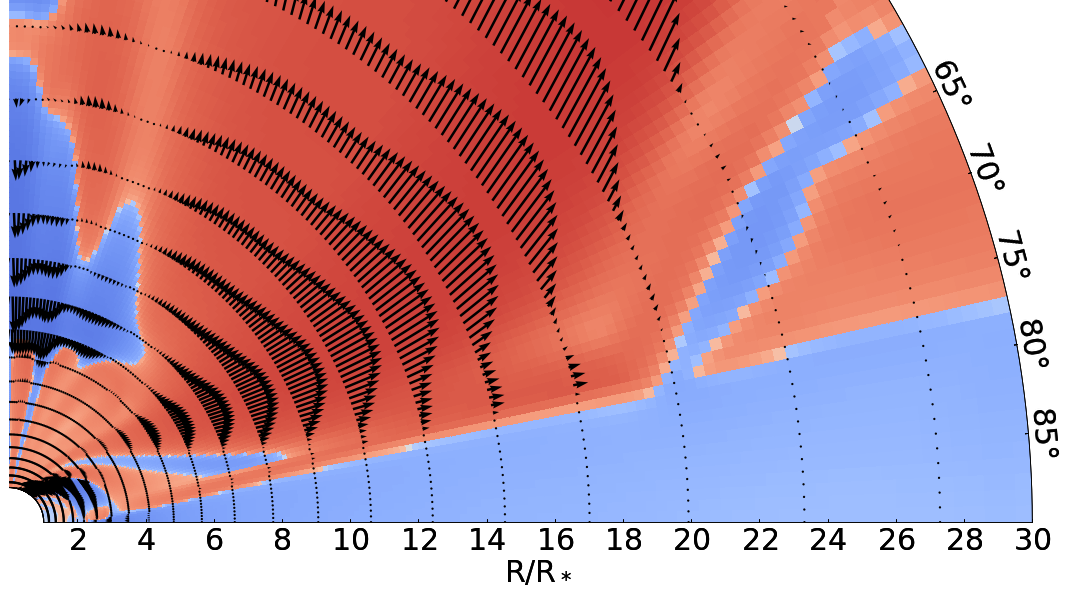}
 \includegraphics[width=\columnwidth]{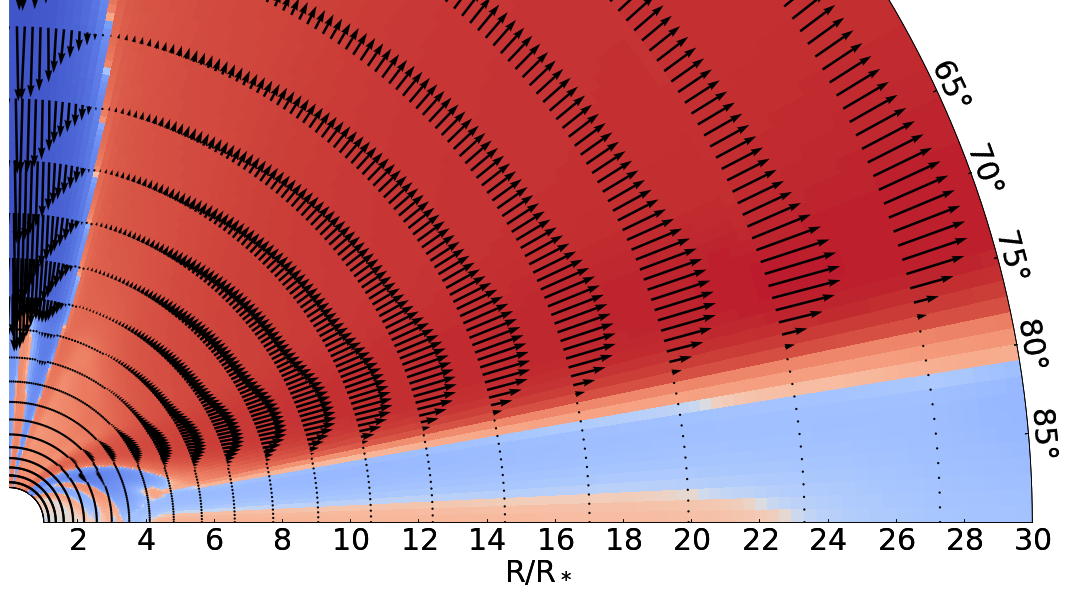}
 \includegraphics[width=\columnwidth]{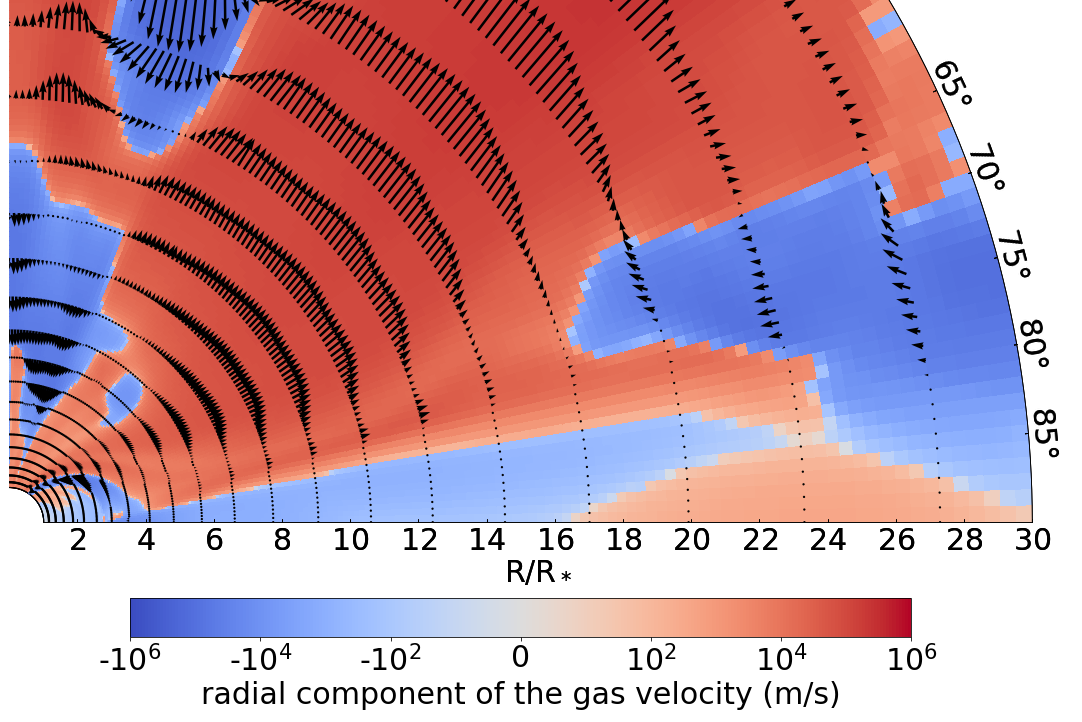}
 \caption{The radial component of the gas velocity (shown as the colour map) and the vectors of gas velocities in our MHD models, from the top to the bottom: model-A, model-B and model-H, respectively. }
 \label{fig:GasVelocity}
\end{figure}

The disc winds exhibit temporal changes on various time scales, even though our solutions are trying to converge to a steady state flow. In the future studies we will explore how dust reacts to this disc wind unsteadiness, especially in the context of sudden puffs of dust lifted out of the disc. In this paper we take snapshots of the wind solution, as shown in Fig.\ref{fig:GasDensity} (gas density structure) and Fig.\ref{fig:GasVelocity} (gas velocity structure), and we follow dust particle trajectories for about an orbital period measured at the orbital radius of initial dust release. This is long enough to see how the dust is impacted by the wind and radiation pressure - whether it stays trapped by the disc gas or it takes on an asymptotic trajectory. 
Our initial discussion of the influence of radiation pressure on the dust particles motion in and above the disc is presented in \cite{Jarosetal20}.

In our MHD simulations, position of the gaseous disc inner rim depends on the stellar rotation rate and strength of the magnetic field. In the cases shown here, it is a footpoint of accretion column (actually an accretion curtain in the axisymmetric 3D case). In the first two cases, accretion flow in the disc is directed towards the star all across the disc height, throughout the whole disc. In the third case, model-H, with $\alpha_{\mathrm v}=0.4$, there is a backflow in the disc \citep{RMishraetal20}. Solutions with a backflow introduce an additional component of the disc flow in the model, which can change the transport of material close to the disc mid-plane, and our inclusion of such solution here is only the first step in the direction of studying such cases.

\section{Optical depth of a dusty disc}
\label{sec:opticaldetph}

The MHD models used in our study do not have dust mixed with gas as a part of the MHD simulations. Since the strength of radiation pressure force depends on the attenuation of stellar irradiation, we devise a method how to postulate the dust optical depth distribution that produces the attenuation. 
Dust in protoplanetary discs is a mixture of grains of various chemical and physical properties, which is further complicated in the inner disc region due to differences in dust sublimation temperatures of different types of grains. This makes dust radiative transfer calculations in optically thick dusty discs extremely complicated and intrinsically unstable \citep{Kama09,Vinkovic12}. 

Fortunately, we are interested in investigating the basic principles of dust dynamics and for that we can exploit two general results of radiative transfer in grain mixtures \citep{Vinkovic06,Kama09,Vinkovic12,Flock16}:
\begin{enumerate}
\item Bigger dust grains are more resilient to sublimation than smaller grains, which results in the "shielding" effect where grains $\gtrsim 1\mu m$ are the closest to the star, while smaller grains coexist with big grains when the (visual) optical depth is $\gtrsim$1 \citep[see Fig.4 in][]{Vinkovic12}.
\item The dust sublimation zone for big grains is not a sharp transition, but rather a large optically thin zone that extends closer to the star and engulfs the entire inner optically thick part of the dusty disc \citep[see Fig.1 in][]{Flock16}.
\end{enumerate}

The first result enables us to postulate a relationship between the optical depth and gas density using the properties of big grains at the location of optically thick inner disc radius, while the theory behind the second result tells us where this radius is located. \citet{Vinkovic06} gives an approximate equation for the disc inner radius:
\begin{equation}
R_{in}=0.0344\Psi \left(\frac{1500 K}{T_{sub}}\right)^2\sqrt[]{\frac{L_{tot}}{L_\odot}}\text{[AU]}
\end{equation}
where $T_{sub}$ is the dust sublimation temperature and $\Psi$ is the correction factor for diffuse heating from the dust itself. The inner radius of optically thick part of dusty disc $R^{thick}_{in}$ has $\Psi\sim 2$. However, as mentioned above, big grains can survive closer to the star under an optically thin regime, where $\Psi\sim 1.2$ gives $R^{thin}_{in}$. This defines the closest possible distance to the star that a big grain can reach. The values of those two critical radii in our models are shown in Table \ref{tbl:Rin}.

\begin{table}
 \caption{Derived stellar luminosities and disc inner radii for the models used in this study.}
 \label{tbl:Rin}
 \begin{tabular}{ccccc}
  \hline
  model & $L_{tot}$ & $\gamma$ & $R^{thick}_{in}$ & $R^{thin}_{in}$ \\
        & $L_\odot$ &          & $R_*$           &  $R_*$ \\
  \hline
  model-A & 0.71    & 0.930 & 6.23 & 3.74 \\
  model-B & 0.668   & 0.988 & 6.05 & 3.63 \\
  model-H & 0.68    & 0.971 & 6.10 & 3.66 \\
  \hline
 \end{tabular}
\end{table}

Note that $R^{thin}_{in}$ is closer to the star than the corotation radius in model-A and model-H ($R_{co}=4.64R_*$) or very close to it ($R_{co}=3.54R_*$ in model-B). Moreover, the most resilient dust grains with $T_{sub}=2000$~K would survive as close as $\sim 2.1R_*$, which is smaller than the truncation radius. However, this part of the disc is exposed to strong viscous and Ohmic heating as the gas moves into the accretion funnel. To prevent too small time step in our MHD simulation, this innermost part of the computational domain, from the foot-point of the accretion column at the disc inner rim to the stellar surface, is treated in ideal-MHD regime, as described in \cite{ZF09}. We do not consider dust under such extreme conditions, where we expect the dust to sublimate due to additional heating sources not accounted for in our model. 

These constraints on the dust sublimation imply that the optical depth, which depends on the dust density, will have three zones: \begin{enumerate}
\item the first is for radii $r<R^{thin}_{in}$ where dust cannot survive and we will use zero optical depth,
\item the second is for radii $R^{thin}_{in}\leq r < R^{thick}_{in}$ where the disc is optically thin, with the dust-to-gas density ratio $\xi^{thin}(r)$, 
\item and the third zone is for $R\geq R^{thick}_{in}$ where the disc is optically thick and we postulate dust-to-gas ratio $\xi^{thick}(r)$. 
\end{enumerate}

The value of $\xi^{thick}$ should be based on a mixing and settling model of the dust within the gaseous disc. Unfortunately, this is a very difficult task, a way beyond the scope of this work and prone to yet another set of various approximations about the dust and gas properties under a highly complex gas density distribution (as seen in Fig.\ref{fig:GasDensity}). Therefore, the common practice is to simply postulate the canonical value of $\xi^{thick}=0.01$ under the assumption of mixing being the same as the primordial dust to gas ratio. We are trying to be a slightly better as we know that this cannot be correct outside of the dense gas regions, where various forces compete to remove the dust as the gas drag is not strong enough to dominate. Thus, we use the following criteria:
\begin{equation}\label{eq:xithick}
  \xi^{thick}=\begin{cases}
    10^{-3}, & \text{if gas density $<5\times 10^{-10}$ kg/m$^3$}.\\
    10^{-2}, & \text{otherwise}.
  \end{cases}
\end{equation}

With $\xi^{thin}$ we resort to a linear approximation that emulates dust-to-gas ratio in the disc model by \citet{Flock16} (their Fig.1)
\begin{equation}\label{eq:xithin}
\log(\xi^{thin}(r))=\log(\xi^{thick}) + \left(7 \frac{\log(r/R^{thin}_{in}) } { \log(R^{thick}_{in}/R^{thin}_{in})} -7\right)
\end{equation}
where the value of $\xi^{thick}$ is based on the local gas density at $\bmath{r}$.
This implies very small amounts of dust in the optically thin disc zone, but the importance of this optically thin zone is in its ability to maintain dust grains that accretion transports through the disc's midplane toward the star. Since this optically thin part of the disc is prone to larger magnetorotational turbulence than the optically thick disc region, dust that enters the optically thin zone might be dispersed to larger disc heights and contributes to the observed near infrared flux \citep{Flock17} or pushed outward back into the disc by the disc wind and radiation pressure. This is also the dust that will experience high temperatures under high gas densities, which could lead to significant physical and chemical alternations of the dust grains. 

Our goal is to find optical depth $d\tau_\lambda$ along the lines that emanate from the star. We start from the basic definition of the optical depth along a path $dl$ 
\begin{equation}\label{eq:taubasic}
d\tau_\lambda=\sigma_\lambda n_{dust}dl
\end{equation}
where $\sigma_\lambda=\pi a^2 Q^{ext}_\lambda$ is the cross section of a dust grain of radius $a$ and the extinction efficiency of $Q^{ext}_\lambda$, while $n_{dust}$ is the local number density of dust grains. We should actually talk about a mixture of dust grains, but we know that only big grains can survive the direct stellar irradiation at these inner disc edges close to the star. Smaller dust appears only when the optical depth becomes large enough to significantly attenuate the stellar spectrum \citep{Vinkovic12}, which also means the attenuation of radiation pressure force. Hence,  the region where the radiation pressure force is important is also the region dominated by big dust grains. Further away from the star smaller grains can also survive the direct stellar radiation, but we keep the big grains as the main contributors to the optical depth in the entire disc surface within our computational domain (i.e. for $r<30 R_*$).

As a canonical big grain we select $a=2\mu$m \citep[the same as in radiative transfer models by][]{Vinkovic12}, which has $Q^{ext}_\lambda\sim 2.5$ at the wavelengths of interest (hence, we can drop $\lambda$ from the equation). The mass of a single dust grain of $\rho_{grain}$ bulk density is $m_d=(4\pi/3)a^3\rho_{grain}$ and the dust number density becomes
\begin{equation}\label{eq:ndust}
n_{dust}=\frac{\rho_{dust}}{m_d}=\frac{3\xi \rho_{gas}}{4\pi \rho_{grain}a^3}
\end{equation}
We combine equation \ref{eq:ndust} with equation \ref{eq:taubasic} and get
\begin{equation}
d\tau_ = \frac{3\xi Q^{ext} \rho_{gas}}{4a\rho_{grain}} dl
\end{equation}
In our case we use $\rho_{grain}=3000 kg/m^3$ and $R_*=2R_\odot$, which gives (recall that our models use $R_*$ as the spatial scale for $dl$)
\begin{equation}\label{eq:tau}
d\tau = 430 \xi\left(\frac{\rho_{gas}}{10^{-9} kg/m^3}\right)\,\, d\left(\frac{l}{R_*}\right)
\end{equation}
where $\xi$ is either $\xi^{thick}$ or $\xi^{thin}$ (equations \ref{eq:xithick} and \ref{eq:xithin}). 
The gas density in our MHD simulations at the midplane results in about an order of magnitude smaller optical depth gradients than typically used in optically thick protoplanetary disc calculations \citep{Vinkovic12}, but it will suffice as far as our approximate estimate of dusty disc winds is concerned. However, if gas densities are such that gas opacity has to be taken into account \citep[e.g. see Section 4.1 in ][]{Dullemond}, this additional contribution is simply added to equation \ref{eq:tau}. 

\section{Radiation pressure force}
\label{sec:RadPress}

The radiation pressure force vector on a dust grain is defined as
\begin{equation}\label{eq:RadPressDef}
\bmath{\mathfrak{F}}=\frac{1}{c}\int \sigma^{ext}_\lambda \bmath{F}_\lambda d\lambda
\end{equation}
where $c$ is the speed of light and $\bmath{F}_\lambda=\bmath{F}^*_\lambda+\bmath{F}^{acc}_\lambda+\bmath{F}^{diff}_\lambda$ is the total radiation flux consisting of the stellar, accretion and diffuse (circumstellar dust and gas) component, respectively. All three flux contributions have details that complicate calculations and require some level of approximation. In this study we deal with the stellar and accretion contributions, while the next paper in the series explores the effects on the dust dynamics due to the diffuse flux component. 

\begin{figure}
 \includegraphics[width=\columnwidth]{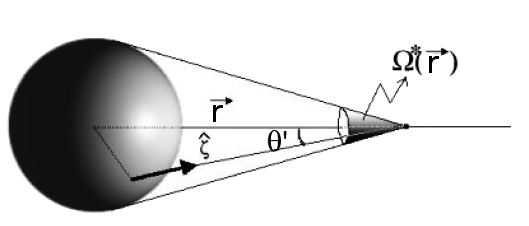}
 \includegraphics[width=\columnwidth]{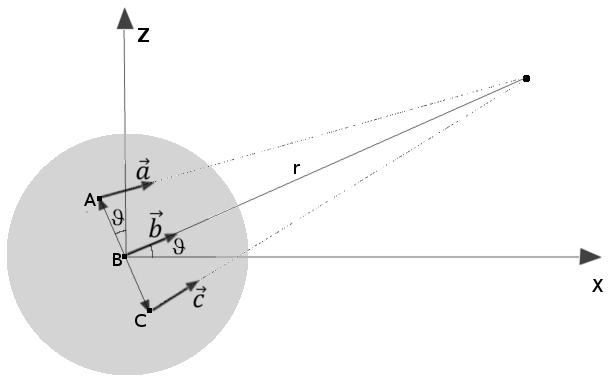}
 \caption{{\it The top panel}: A dust grain at position $\bmath{r}$ is illuminated by stellar radiation coming into the grain within the solid angle $\Omega^*(\bmath{r})$. Rays between the star and the grain are defined by the unit vector $\hat{\zeta}$ that closes an angle $\theta^\prime$ with the vector $\bmath{r}$. {\it The bottom panel}:  A complicated integral in equation \ref{eq:Fstarinteg} is replaced with contributions from three directions $\bmath{a}$,$\bmath{b}$ and $\bmath{c}$ as shown in this figure. Each direction points toward the dust grain from the middle of its third of the stellar surface, marked with $A$, $B$ and $C$.  A dust grain is positioned at distance $r$ and angle $\vartheta$ relative to the disc midplane.}
 \label{fig:StarSurface}
\end{figure}

\subsection{Stellar and accretion flux contribution}

The stellar and accretion contribution, summed together as $\bmath{F}^{star}_\lambda=\bmath{F}^*_\lambda+\bmath{F}^{acc}_\lambda$, require integration of the radiation intensity over the emitting surface visible from a dust grain located at $\bmath{r}$. We express this with the equation
\begin{equation}
\bmath{F}^{star}_\lambda(\bmath{r})=\int  \left(I^*_\lambda(\hat{\zeta}) + I^{acc}_\lambda(\hat{\zeta})\right)\, e^{-\tau_\lambda(\hat{\zeta})} \cos \theta^\prime d\Omega^*(\hat{\zeta})\hat{\zeta}
\end{equation}
where $I^*_\lambda(\hat{\zeta})$ and $I^{acc}_\lambda(\hat{\zeta})$ are the stellar photosphere radiation and the accretion shock intensities emitted from the stellar surface into the direction $\hat{\zeta}$, $\tau_\lambda(\hat{\zeta})$ is the optical depth along the path between the stellar surface and the dust grain, $\theta^\prime$ is the angle between $\hat{\zeta}$ and radial direction $\bmath{r}$, and $d\Omega^*(\hat{\zeta})$ is the solid angle covered by a small portion of the emitting stellar surface. See the top panel in Fig.\ref{fig:StarSurface} for a sketch of this geometry.

In reality this is a highly complicated integral because the stellar surface has a complicated brightness structure, especially with the accretion shock included into the picture, and the complexity of optical depths introduced by the dust and gas structure of accretion disc. Hence, we need to resort to approximations that will give us a general insight into the importance of radiation pressure to the dust dynamics. We will approximate the accretion shock radiation propagating only along the radial direction $\bmath{r}$, while the stellar photosphere intensity will be angle independent and equal to the Planck function $I^*_\lambda(\hat{\zeta})=B_\lambda(T_*)$. This gives
\begin{equation}\label{eq:Fstarinteg}
\begin{split}
\bmath{F}^{star}_\lambda(\bmath{r})=\frac{L_*}{4\pi r^2}b_\lambda(T_*)\int  e^{-\tau_\lambda(\hat{\zeta})} \cos \theta^\prime \frac{d\Omega^*(\hat{\zeta})}{\Omega^*}\hat{\zeta}  +\\
+\frac{L_{acc}}{4\pi r^2}b_\lambda(T_{acc})e^{-\tau_\lambda(\hat{r})}\hat{r}
\end{split}
\end{equation}
where we use the luminosity and distance $r$ to express the bolometric flux and the total solid angle is $\Omega^*=\pi (R_*/r)^2$. The optical depths can be calculated using equation \ref{eq:tau}, but we need to simplify the integral over the stellar surface. We do that by taking only three directions, each representing one third of the stellar disc (see the bottom panel in Fig.\ref{fig:StarSurface}). We combine this with the spectrum approximations in equation \ref{eq:fstar} and get
\begin{equation}\label{eq:flux}
\bmath{F}^{star}_\lambda(\bmath{r})=\frac{L_{tot}}{4\pi r^2}\left(\gamma b_\lambda(T_*)
\bmath{\varepsilon} + (1-\gamma)b_\lambda(T_{acc})e^{-\tau(\hat{r})}\,\,\hat{r} \right)
\end{equation} 
\begin{equation}\label{eq:f}
\bmath{\varepsilon}\approx \sum_{\hat{\zeta}=\hat{a},\hat{b},\hat{c}}e^{-\tau(\hat{\zeta})}\,\,
\frac{\hat{\zeta}}{3} 
\end{equation}
where the unit vectors $\hat{a}$, $\hat{b}$ and $\hat{c}$ are normalized vectors
\begin{equation}\label{eq:abc}
\bmath{a}=\bmath{r}-h(-\sin\vartheta,\cos\vartheta); \,\,\,\,\,
\bmath{b}=\bmath{r}\,; \,\,\,\,\,
\bmath{c}=\bmath{r}-h(\sin\vartheta,-\cos\vartheta)
\end{equation}
where $h$=0.553 and $\vartheta$ is the angle between $\bmath{r}$ and the disc midplane (Fig.\ref{fig:StarSurface}). 

Notice how the vector $\bmath{\varepsilon}$ is non-radial at the disc surface and points slightly into the disc interior because the optical depth along $\hat{c}$ is larger than the depth along $\hat{a}$. Examples of the spatial distribution of $|\bmath{\varepsilon}|$ are shown in Fig.\ref{fig:GasDensity} as contours. 

\subsection{Radiation pressure vector}  

The flux from equation \ref{eq:flux} can be now used in the radiation pressure equation \ref{eq:RadPressDef} to obtain the force vector $\bmath{\mathfrak{F}}^{star}$. We scale this with the local gravity force to derive the "strength" of the radiation pressure force on a dust grain of mass $m_d$ and radius $a$
\begin{equation}\label{eq:betaqext}
\bmath{\beta}=\frac{\bmath{\mathfrak{F}}^{star}}{GM_*m_d/r^2}=\frac{L_{tot}a^2}{4cGM_*m_d}\bmath{q}_{ext}
\end{equation}
\begin{equation}\label{eq:qext}
\bmath{q}_{ext}=\gamma \langle Q^{ext}\rangle_{T_*}\bmath{\varepsilon} + (1-\gamma)\langle Q^{ext}\rangle_{T_{acc}}e^{-\tau}\,\,\hat{r}
\end{equation}
where $G$ is the gravity constant, and $\langle Q^{ext}\rangle_T$ is the Planck average of the extinction coefficient defined as
\begin{equation}
\langle Q^{ext}\rangle_{T} = \int Q^{ext}_\lambda b_\lambda(T)  d\lambda
\end{equation}
We rewrite equation \ref{eq:betaqext} into a more convenient form
\begin{equation}\label{eq:beta}
\bmath{\beta}=0.19 \left(\frac{L_{tot}}{L_\odot}\right)\left(\frac{M_\odot}{M_*}\right)
\left(\frac{3000 kg/m^3}{\rho_{grain}}\right)\left(\frac{\mu m}{a}\right) \bmath{q}_{ext}
\end{equation}
where $\rho_{grain}$ is the grain's bulk density. This equation is used in our calculations of the radiation pressure force. 

\begin{figure}
 \includegraphics[width=\columnwidth]{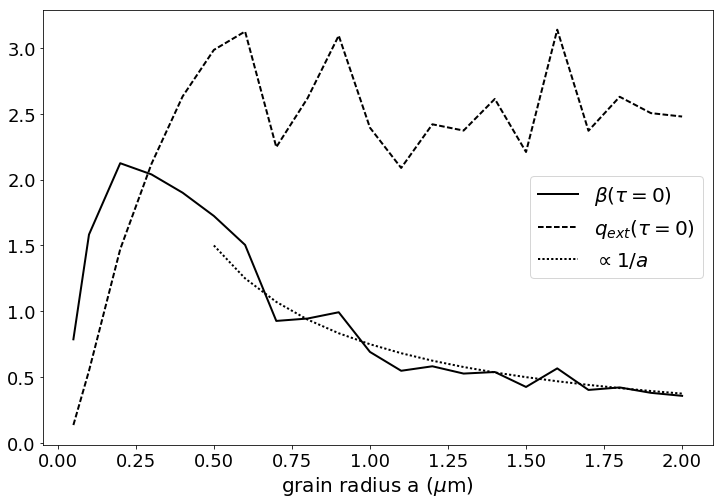}
 \caption{The strength of radiation pressure force $\beta$ (equation \ref{eq:beta}) as a function of the grain radius $a$ in the case of no stellar obscuration (optical depth $\tau=0$). The dust optical properties are taken from \citet{Dorschner95} for olivine grains. The stellar properties are from the model-A. The radiation pressure increases steeply with the grain size for small grains, reaches the peak at grains of 0.25-0.5$\mu$m in radius and then declines as $\propto a^{-1}$ (the dotted line) for $a\gtrsim 0.5\mu m$. The decline results from the dust extinction being a constant at the wavelengths of interest. This is also visible in $q_{ext}\sim 2.5$ for grain radii $a\gtrsim 0.5\mu m$ (the dashed line). The range of dust grain sizes prone to ejection (i.e. $\beta>0.5$) differs with the stellar types (changes in luminosity and spectral shape) and grain porosity \citep{WolfBeta}.
 }
 \label{fig:beta}
\end{figure}

When a dust grain is released from the disc, its initial velocity is close to the Keplerian. This means that grains with $\beta>0.5$ will be ejected outward \citep{KrivovBeta}, which results in a grain leaving the system or re-entering the disc at larger distances. Although in the next sections we combine the radiation pressure with the gas drag, it is informative to look at the value of $\beta$ when the optical depth is zero ($\tau=0$). This represents the maximum strength of  radiation pressure, when a dust grain is not obscured by other dust within the disc. 
The tricky part in calculating $\bmath{\beta}$ in our case is hidden in $\bmath{q}_{ext}$ (equation \ref{eq:qext}). But, in the case of $\tau=0$, its vector direction is purely radial and the magnitude is simply
\begin{equation}\label{eq:Qstar}
q_{ext}(\tau=0)= \int Q^{ext}_\lambda f^{star}_\lambda  d\lambda
\end{equation}

Fig.\ref{fig:beta} shows the dependence of $\beta(\tau=0)$ on the grain radius. In our calculations we use the optical properties of olivine dust grains \citep{Dorschner95} and the bulk density of $\rho_{grain}=3000 kg/m^3$.  For very small grains ($\lesssim 0.1\mu m$) the flux from T Tau stars is too low to push the grains. However, this can be changed if the accretion is stronger and contributes more to the UV part of the spectrum. In our example, the radiation pressure peaks at $\beta\sim 2$ for grains of about $\sim 0.5\mu m$ in size (twice the radius) and then drops as $a^{-1}$ (the geometrical optics regime with the constant value of $q_{ext}\sim 2.5$). The largest grain radius still to be blown out in our T Tau examples (i.e. with $\beta> 0.5$) is $a \sim 2\mu m$. Notice that $\beta$ scales with the stellar luminosity (equation \ref{eq:beta}). This means that in the case of Herbig Ae stars, where the luminosities are an order of magnitude or more larger than in our T Tau examples, radiation pressure is so strong that it can reshape the dusty disc surface \citep{Vinkovic14} and easily produce large scale mixing in protoplanetary discs \citep{Vinkovic09}.

\subsection{Porous grains}

Modelling the dust dynamics only with compact grains of no porosity is probably oversimplistic. Recent in-situ discoveries of highly porous grains in cometary dust (with bulk density $\lesssim$1000~kg/m$^3$) \citep{RosettaFulle,RosettaMannel} implies that radiation pressure could play a more prominent role than initially considered because porous grains experience much stronger radiation pressure force than non-porous grains of a similar size \citep{Tazaki}. Dust grains can stick together into agglomerates, which then do not follow the mass-to-size $m_d\propto a^{3}$ relationship, but rather some fractal dimension $D_f$ due to porosity $m_d\propto a^{D_f}$. For example, in-situ measurements of dust in comet 67P/Churyumov--Gerasimenko \citep{RosettaFulle, RosettaMannel, Mannel19, Bentley}  showed that the grains have $D_f\sim 1.7$, with the bulk density $\lesssim$1000~kg/$m^3$, and they consists of sub-unites of 1-2$\mu m$ in size. The size of sub-unites agrees with the range of grain sizes in Fig.\ref{fig:beta} that can be pushed outward by the radiation pressure in T Tau stars. 

Porosity with a low $D_f$ significantly alters $\beta$. \citet{WolfBeta} explore this effect for different stellar types (an advanced version of our Fig.\ref{fig:beta}).  Big porous grains have ${q}_{ext}\sim 2.5$, but from equation \ref{eq:betaqext} we see that $\beta\propto a^2/m_d$. While compact grains have $m_d\propto a^{3}$, which results in $\beta\propto a^{-1}$, the porous grains of $D_f\sim 2$ have $\beta \approx const$. This means that $\beta$ in porous grains is similar to the value of their sub-units, even though the grain agglomerates are much bigger than the sub-units \citep{Mukai92,Tazaki}. Since our goal is to see some general solutions to the dust dynamics, and not to explore fine details of the dust properties, we address porosity by simply using a lower dust bulk density of  1000$kg/m^3$ in equation \ref{eq:beta}. This is  equivalent to increasing the grain porosity. 

\section{Dust motion under gas drag and radiation pressure}
\label{sec:equations}

\subsection{Gas drag}

Gas drag force on dust particles applicable in our case is described with the Epstein regime, where the drag force per particle mass $m_d$ is
\begin{equation}\label{eq:dragfroce}
\frac{\bmath{F}_{drag}}{m_d}=\frac{\rho_{gas}}{\rho_{grain}}\,\,\frac{\text{v}_{th}}{a}(\bmath{V}_{grain}-\bmath{V}_{gas})
\end{equation}
where $\rho_{gas}$ is the gas density, $\rho_{grain}$ is the grain's bulk density, $\text{v}_{th}=\sqrt{8kT/\pi\langle m_g\rangle}$ is the average thermal speed of the gas molecules of the average molecular mass $\langle m_g\rangle=2.3 m_H$ ($m_H$ is the mass of hydrogen atom) at temperature $T$, $a$ is the grain's radius, $\bmath{V}_{grain}$ is the grain's velocity and $\bmath{V}_{gas}$ is the local velocity of gas. 

A useful measure of the importance of gas drag is the stopping timescale (or drag timescale) $t_s$. It is an estimate of the time required by the frictional drag to cause an order-of-unity change in the momentum of the dust grain
\begin{equation}\label{eq:ts}
t_s=\frac{m_d|\bmath{V}_{grain}-\bmath{V}_{gas}|}{|\bmath{F}_{drag}|}=\frac{\rho_{grain}}{\rho_{gas}}\,\,\frac{a}{\text{v}_{th}}
\end{equation}
This timescale can be compared with the local orbital (Kepler) period
$t_K={2\pi}/{\Omega_K}=2\pi t_0 ({r}/{R_*})^{3/2}$. 
Dust grains are dominantly decoupled from the gas motion if $t_s>t_K$, and strongly influenced by the gas velocity when $t_s<t_K$.

Without the influence of radiation pressure these timescales would give us an informed guess on the dynamics of dust grains. However, the introduction of radiation pressure force can significantly alter such an analysis, as we show in the next section. Even if the radiation pressure acts as a small perturbation, it should not be ignored because it can alter dust particle trajectories and influence the dust mixing and settling processes in the disc. 

\subsection{Dust and gas temperature}
\label{sec:temperature}

Our MHD simulations do not yield the gas temperature, which means we have to introduce some other way of calculating it. This is a highly complex task, but we can again exploit some general physical properties of dust and gas to simplify the procedure. First, we are not interested in the optically thick disc interior (where the dust radiative transfer is very difficult to calculate) because the radiation pressure becomes negligible and ordinary gas drag calculations apply. Second, the dust and gas temperatures are decoupled in the region outside the optically thick dusty disc and their temperatures should be determined separately. However, we need the gas temperature only to calculate the sound speed, where the temperature enters the gas drag equation \ref{eq:dragfroce} with the square root and shows small spatial gradients, while the gas density enters linearly and shows large spatial gradients (Fig.\ref{fig:GasDensity}). Therefore, dust trajectories will be dominated by the gas gradients, while the square root temperature errors are a small perturbance as long as we are not a way off in the temperature estimate.  Numerical models \citep[e.g.][]{Woitke} indicate that only small regions just above the inner dusty disc rim might have the most extreme gas to dust temperature difference that would increase the sound speed by about 50\%. This means the gas drag would be increased by the same factor and help the gas wind to lift the dust. However, this is also the zone of large gas density gradient, where such a correction on the sound speed would not significantly reshape the overall size of gas drag region. 

Thus, we will approximate the gas temperature with the dust temperature, which can be off by no more than a factor of a few in the worst case scenarios. We calculate the dust temperature by ignoring the diffuse contribution, which is an acceptable approach because we care only about the dust in the disc surface or above the dusty disc, where the dust heating is dominated by the stellar radiation. Under the local thermodynamic equilibrium the heating of a dust grain of absorption coefficient $Q^{abs}_\lambda$ is equal to the grain's thermal emission
\begin{equation}
\int J^{tot}_\lambda\, Q^{abs}_\lambda d\lambda = \int B_\lambda(T_{dust})\,\, Q^{abs}_\lambda d\lambda
\end{equation}
where $J^{tot}_\lambda$ is the radiative intensity from the star (and includes both the stellar and the accretion shock radiation). We proceed in the same way as in the case of the flux equation \ref{eq:flux} and obtain
\begin{equation}\label{eq_Tdust}
T^4_{dust} = \frac{T^4_{*}}{4\gamma (r/R_*)^2}\,\,\frac{\gamma \langle Q^{abs}\rangle_{T_*}|\bmath{\varepsilon}|+(1-\gamma)\langle Q^{abs}\rangle_{T_{acc}}e^{-\tau}}{\langle Q^{abs}\rangle_{T_{dust}}}
\end{equation}
where we used $L_{tot}=L_*/\gamma$.

Notice that the dust grain temperature appears on both sides of equation \ref{eq_Tdust}, which means that $T_{dust}$ needs to be extracted through iterations. The first iteration uses $T_{dust}=T_{sub}$ everywhere and then only a few iterations are needed to achieve acceptable convergence. Although our approximation does not include the diffuse heating from the dust itself, such a correction impacts our estimate of the sound speed by less than 10\%. This is a negligible effect on the level of our general exploration of possible scenarios regarding dust trajectories.

\subsection{Equations of motion}

Our equation of motion for a single dust grain at position $\bmath{r}$ includes forces of gravity, gas drag and radiation pressure \citep{Takeuchi,Vinkovic09}
\begin{equation}\label{eq:ForceEquation}
\ddot{\bmath{r}}=-G\frac{M_*}{r^3}\,\bmath{r}-\frac{\rho_{gas}}{\rho_{grain}}\,\,\frac{\text{v}_{th}}{a}(\dot{\bmath{r}}-\bmath{V}_{gas})
+G\frac{M_*}{r^2}\,\bmath{\beta}
\end{equation}
We now scale this equation with the normalization units used in the MHD simulation: $R_*$ as the unit length and $t_0=(R_*^3/GM_*)^{1/2}$ as the unit time. The new scaled variables are $r\equiv r/R_*$, $t\equiv t/t_0$ and $\bmath{V}\equiv \bmath{V}\,t_0/R_*$, which leads to
\begin{equation}\label{eq:ForceEquationScaled}
\ddot{\bmath{r}}=-\frac{\bmath{r}}{r^3}-\frac{\rho_{gas}\,\text{v}_{th}\,t_0}{\rho_{grain}\,a}(\dot{\bmath{r}}-\bmath{V}_{gas})
+\frac{\bmath{\beta}}{r^2}
\end{equation}
The equation can be simplified if we introduce a dimensionless parameter
\begin{equation}\label{eq:muEquation}
\mu(\bmath{r})=\frac{\rho_{gas}\,\text{v}_{th}\,t_0}{\rho_{grain}\,a}
\end{equation}
which becomes
\begin{equation}\label{eq:mu}
\mu(\bmath{r})=\frac{2\times 10^9 \rho_{gas}(\bmath{r}) }{a}
\left(\frac{R_*}{R_\odot}\right)^{3/2}
\left(\frac{M_\odot}{M_*}\right)^{1/2}
\left(\frac{T(\bmath{r})}{1500}\right)^{1/2}
\left(\frac{3000}{\rho_{grain}}\right) 
\end{equation}
where $\rho_{gas}$ and $\rho_{grain}$ are in the units of $kg/m^3$ and $a$ in $\mu m$.

We break down this equation into three components using the cylindrical coordinate system $\bmath{r}=(\varrho,\varphi,z)$ and $\bmath{V}_{gas}=(\text{v}_{gas,\varrho},\text{v}_{gas,\varphi},\text{v}_{gas,z})$:
\begin{equation}\label{eq:ForceRho}
\ddot{\varrho}=\varrho \dot{\varphi}^2 - \frac{\varrho}{(\varrho^2+z^2)^{3/2}} - \mu (\dot{\varrho}-\text{v}_{gas,\varrho})+\frac{\beta_\varrho}{\varrho^2+z^2}
\end{equation}
\begin{equation}\label{eq:ForcePhi}
\ddot{\varphi}=-2\frac{\dot{\varrho}}{\varrho}\dot{\varphi}-\mu \left(\dot{\varphi}-\frac{\text{v}_{gas,\varphi}}{\varrho}\right)
\end{equation}
\begin{equation}\label{eq:ForceZ}
\ddot{z}= - \frac{z}{(\varrho^2+z^2)^{3/2}} - \mu (\dot{z}-\text{v}_{gas,z})+\frac{\beta_z}{\varrho^2+z^2}
\end{equation}
where the radiation pressure vector from equation \ref{eq:betaqext} is axially symmetric: $\bmath{\beta}=(\beta_\varrho,0,\beta_z)$. 
We integrate these three equations to reconstruct a dust grain trajectory. 

\subsection{Radiation pressure relative to the drag force}

The importance of radiation pressure relative to the gas drag can be deduced by comparing $\beta$ with $\mu$. Both of those parameters depend on the spatial coordinates, thus we can create a map of their ratio to see where one dominates over the other:
\begin{equation}\label{eq:D}
\bmath{\mathfrak{D}}=\frac{\bmath{\beta}}{\mu}=
\frac{10^{-10}}{\rho_{gas}}
\left(\frac{R_\odot}{R_*}\right)^{3/2}
\left(\frac{L_*}{L_\odot}\right)
\left(\frac{M_\odot}{M_*}\right)^{1/2}
\left(\frac{1500}{T}\right)^{1/2}
\bmath{q}_{ext}
\end{equation}
were $\rho_{gas}$ is in the units of $kg/m^3$.

Notice that this ratio does not depend neither on the dust grain bulk density nor on the grain radius. The main contributor to the changes in $\mathfrak{D}$ is the gas density. 
The zone of $\mathfrak{D}\sim 1$ maps the dynamic's shift from the dominance of gas drag when $\mathfrak{D}<1$ to the dominance of radiation pressure force when $\mathfrak{D}>1$. The zone where radiation pressure is not dominant, but still not negligible ($\mathfrak{D}\lesssim1$) is also interesting to study, as we will see in the results below. 

\subsection{The initial dust velocity}

When we start trajectory integration to follow the dynamics of a dust grain, we need to specify the initial grain velocity. Ideally, we would like to see a dust grain in a steady state flow under the influence of all the forces. Equation \ref{eq:D} shows two regimes as a competition between the gas drag and the radiation pressure. The steady state solution under the gas drag dominance is
\begin{equation}\label{eq:DustVelRho0}
\dot{\varrho}(t=0)\sim \text{v}_{gas,\varrho}+\frac{1}{\mu}
\left(\varrho \dot{\varphi}^2 - \frac{\varrho}{(\varrho^2+z^2)^{3/2}} +\frac{\beta_\varrho}{\varrho^2+z^2}\right)
\end{equation}
\begin{equation}\label{eq:DustVelPhi0}
\dot{\varphi}(t=0)\sim \frac{\text{v}_{gas,\varphi}}{\varrho}
\end{equation}
\begin{equation}\label{eq:DustVelZ0}
\dot{z}(t=0)= \text{v}_{gas,z}+\frac{1}{\mu}
\left( - \frac{z}{(\varrho^2+z^2)^{3/2}} +\frac{\beta_z}{\varrho^2+z^2}\right)
\end{equation}

A problem arises in the zone of radiation pressure dominance. The steady state for a grain experiencing mostly the radiation pressure force is an orbit that is either an ellipse at some larger distance from the star, as the gravity is reduced by the radiation pressure force, or an open trajectory out of the system. Such a steady state orbit is typically not at the location of the initial dust grain position when the integration starts. This means that we can not simply drop grains into the $\mathfrak{D}> 1$ zone and start integrating, except for the cases when $\mathfrak{D}\sim 1$, where the above steady state equations \ref{eq:DustVelRho0}-\ref{eq:DustVelZ0} still do not diverge. Dust dynamics in $\mathfrak{D}> 1$ can be integrated only if the initial dust position was set into the region where $\mathfrak{D}\lesssim 1$ 

Another interesting dust dynamics behaviour is revealed if we consider how the steady state solution changes with the dust properties. For example, reducing the grain size $a$ increases $\mu$ by the factor of $1/a$. At the same time, $\mathfrak{D}$ changes only slightly as it depends on $|\bmath{q}_{ext}|$, which changes slowly with $a$ when the radiation pressure is not negligible. If we apply these changes to the steady state equations, we can see that the gas drag would push the dust speed closer to the local gas speed, but the radiation pressure component of the equations remains almost unchanged. In other words, the relative importance of radiation pressure for the dust grain trajectory increases with a decreasing $a$ when the radiation pressure is not negligible. Another example is fluffy grains, where we reduce $\rho_{grain}$, which increases $\mu$ but has no influence on $\mathfrak{D}$. Hence, the dynamics of fluffy grains is influenced more strongly by the radiation pressure force compared to the same grains sizes of ordinary solid grains, even though the drag force on fluffy grains is increased. 

\section{Results}
\label{sec:results}

\subsection{The algorithm layout}
\label{subsec:algorithm}

We are interested in mapping possible local dust movements within our quasi-stationary MHD models. The focus is on dust particles in the disc surface, which means we need only short term integration of dust dynamics. This is performed with the Leapfrog integration of dust dynamics equations \ref{eq:ForceRho}-\ref{eq:ForceZ} during no more than a few dust orbital periods (measured at the radius of initial dust release). Here we explore how much radiation pressure perturbs the projected displacement of dust particles compared to the gas drag only. A more complex long-duration integration, that takes into account the time-varying component of the quasi-stationary MHD solutions, is left for the future work. 

The dust grains in our modelling are olivine silicates from \citet{Dorschner95} with two bulk densities: 3000~kg/m$^3$ for compact grains and 1000~kg/m$^3$ for porous grains. The optical depth and temperature structure are calculated with $a=2\mu$m grains (see sections \ref{sec:opticaldetph} and \ref{sec:temperature}) and then used as input parameters for the dust dynamics integrator. 
The sequence of computational steps is as follows:
\begin{enumerate}
\item Set the grain size $a$ and import its optical properties $Q^{ext}_\lambda$ and $Q^{abs}_\lambda$.
\item Import gas density $\rho_{gas}$ and velocity $\bmath{V}_{gas}$ from the MHD simulation.
\item Calculate an array of optical depths on the MHD spatial grid using equations \ref{eq:tau} and \ref{eq:xithin}. 
\item Calculate an array of $\bmath{\varepsilon}$ using equation \ref{eq:f}.
\item Calculate an array of $\bmath{q}_{ext}$  using equation \ref{eq:qext}.
\item Calculate an array of $\bmath{\beta}$ using equation \ref{eq:beta}.
\item Calculate the temperature array using equation \ref{eq_Tdust}.
\item Calculate an array of $\mu$ using equation \ref{eq:mu}.
\item Define the initial position of a dust grain and set the initial velocity using the steady state equations \ref{eq:DustVelRho0}-\ref{eq:DustVelZ0} (be aware of the constraints when $\mathfrak{D}> 1$).
\item Integrate equations \ref{eq:ForceRho}-\ref{eq:ForceZ} to follow the dust grain trajectory. 
\end{enumerate}

We position dust grains on various locations in the disc surface zone in a way that reveals how the dust would flow. Then we run dust trajectory calculations and visualise dust paths in the $\varrho-z$ plane. We also run calculations with the radiation pressure force turned off to compare the trajectories with the same calculations when the radiation pressure is included. This comparison gives us a sense of how strong the radiation pressure effect is and what impact it has on the trajectory shape. The grain radii representative of possible trajectory outcomes in our models are 2$\mu$m, 0.2$\mu$m and 0.1$\mu$m. We present two views of the disc: one up to 30 stellar radii and the other up to 10 stellar radii. The former view is used for exploring the overall disc surface dust behaviour, while the latter is focused on the dust flow in the zone of inner dusty disc edge between $R^{thick}_{in}$ and $R^{thin}_{in}$. We do not consider $a=0.1\mu$m grains in this edge zone because it is too hot for those grains to survive, although it is not excluded that they  occasionally appear after breakup of fluffy grains during grain collisions. 

\begin{figure}
 \includegraphics[width=\columnwidth]{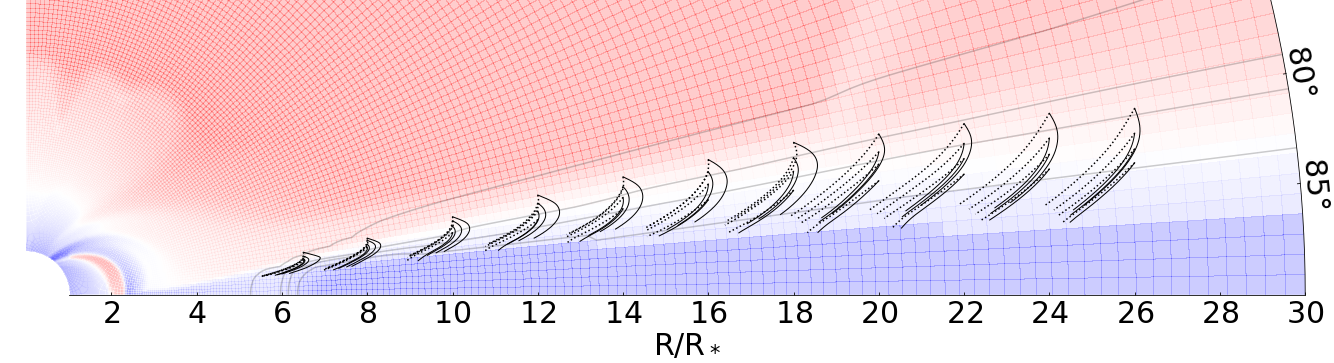}
 \includegraphics[width=\columnwidth]{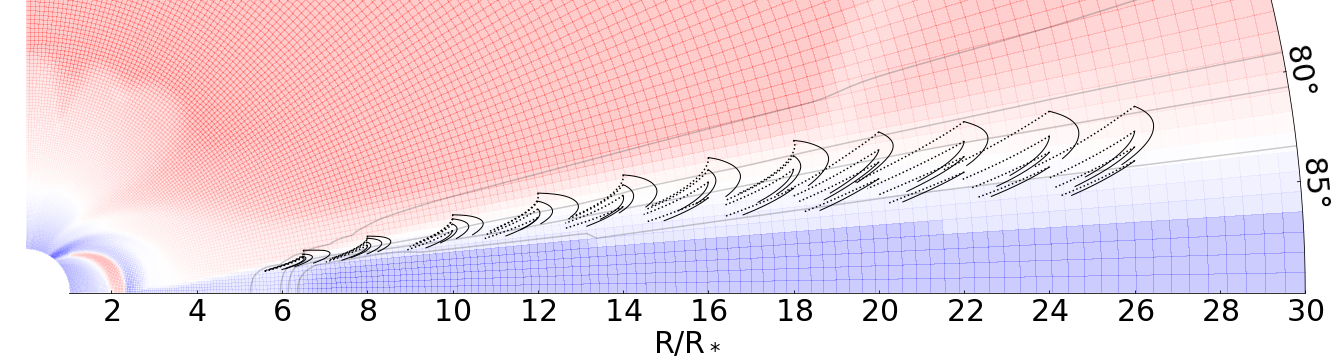}
 \includegraphics[width=\columnwidth]{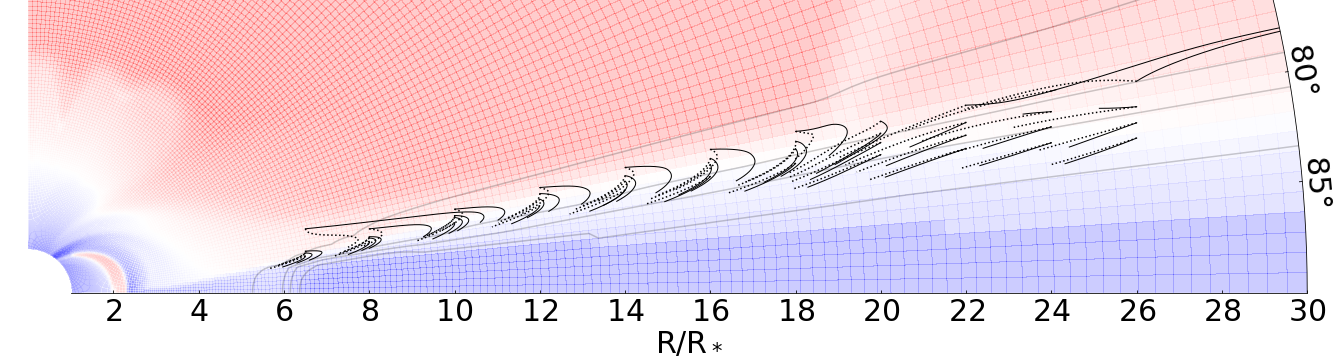}
 \includegraphics[width=\columnwidth]{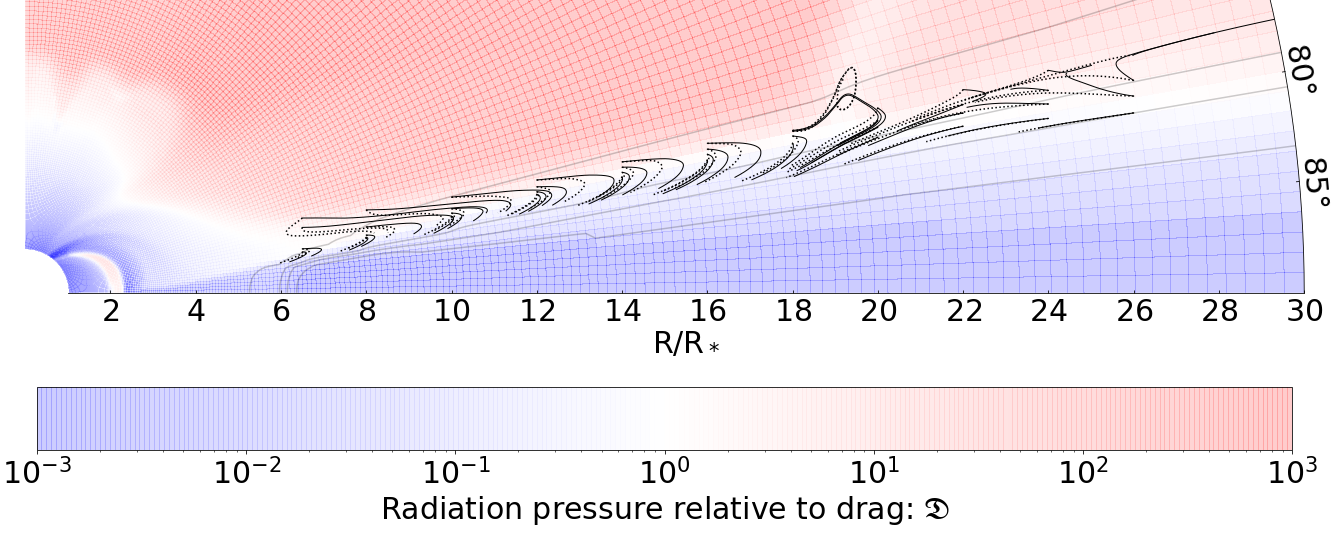}
 \caption{Trajectories of dust grains at different locations over the disc surface in the outflow model-A. See Fig. \ref{fig:GasVelocity} for a view of the gas velocity flow. The colour map shows the strength of radiation pressure relative to the gas drag ($\mathfrak{D}$, equation \ref{eq:D}). The thin solid lines are contours of the optical depth attenuation factor $\varepsilon$ described by equation \ref{eq:f}, with values set to 0.1, 0.4, 0.7 and 0.99 (the maximum is 1.0 in regions without dust). The thick solid lines are paths of dust particles with the radiation pressure force included into the dynamics, while the dotted lines are the same particles with the radiation pressure turned off for comparison. The bulk density of dust grains is 3000 kg/m$^3$, except in the second from the top panel where it is 1000 kg/m$^3$. The grain radius differs between the panels: 2$\mu$m in the top two panels, 0.2$\mu$m in the third panel and 0.1$\mu$m }in the bottom panel. The dust flow is in all cases outward and/or into the disc. 
 \label{fig:modelAsurfaceFlow}
\end{figure}

\subsection{The disc surface flow}

\subsubsection{The model-A flow}

The MHD model-A displays a complicated behaviour of the gas inflow and outflow over the disc surface (see Fig.\ref{fig:GasVelocity}). Hence, we expect that some dust particles will be dragged along by the gas into equally complicated trajectories. Under such conditions, the fate of a dust particle depends on the interplay between the particle's properties and various physical conditions at the particle's location, mainly the gas density and the strength of radiation pressure relative to the gas drag ($\mathfrak{D}$). Fig.\ref{fig:modelAsurfaceFlow} shows results for our test particles initially distributed within the disc surface zone of $\mathfrak{D}\sim 1$. 

The $a=2\mu$m grains quickly sink into the disc due to the gas drag, but a comparison with the trajectories that have radiation pressure turned off reveals how the radiation pressure enables the grains in $\mathfrak{D}\gtrsim 1$ to move slightly outward before the drag completely starts to dominate deeper within the disc. Without radiation pressure grains display only inward motion. This effect is much more pronounced in fluffy grains (i.e. grains with the bulk density of 1000 kg/m$^3$ in our models), which have a stronger outflow trajectory distortion due to the radiation pressure and a much faster inflow trajectory when this pressure is turned off. This behaviour is essentially what \citet{Takeuchi} described for the surface of optically thick discs where the radiation pressure force creates a net dust outflow and contributes to the local dust mixing. 

With smaller grains of $a=0.2\mu$m and $0.1\mu$m we see that the effects of radiation pressure are more pronounced. There is a sharp transition between the gas drag dominated region of $\mathfrak{D}<1$, where the trajectories with and without radiation pressure are quite similar, and the region of $\mathfrak{D}>1$, where grains experience a strong outflow drift. In the extreme cases, the grains are pushed far away by the radiation pressure force. Moreover, complex trajectory shapes emerge in the region of gas wind velocity reversal, where at $\varrho\sim 20$ a wall of gas density above the disc (see Fig.\ref{fig:GasDensity}) flows toward the star instead outward as the rest of the wind (see Fig.\ref{fig:GasVelocity}). This MHD effect on the gas flow stirs up the disc surface dust and enables it to fly far away toward the outer parts of the disc. 

\begin{figure}
 \includegraphics[width=\columnwidth]{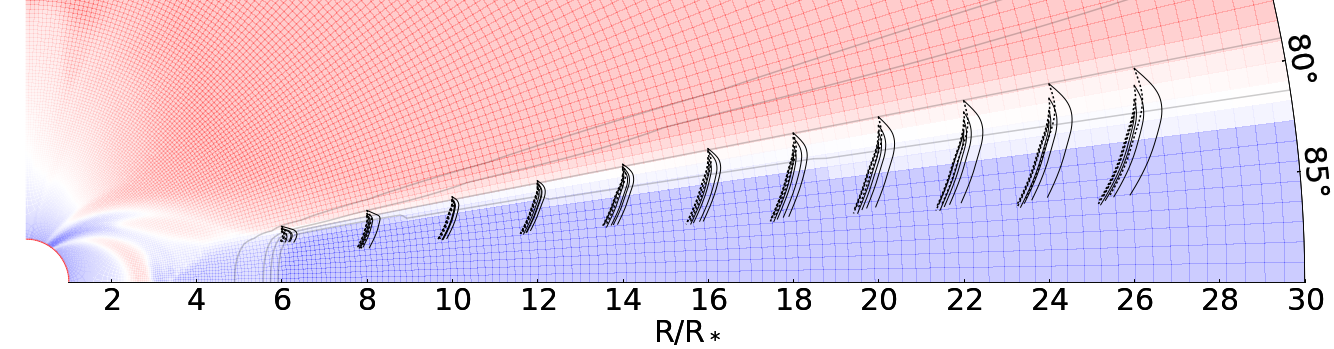}
 \includegraphics[width=\columnwidth]{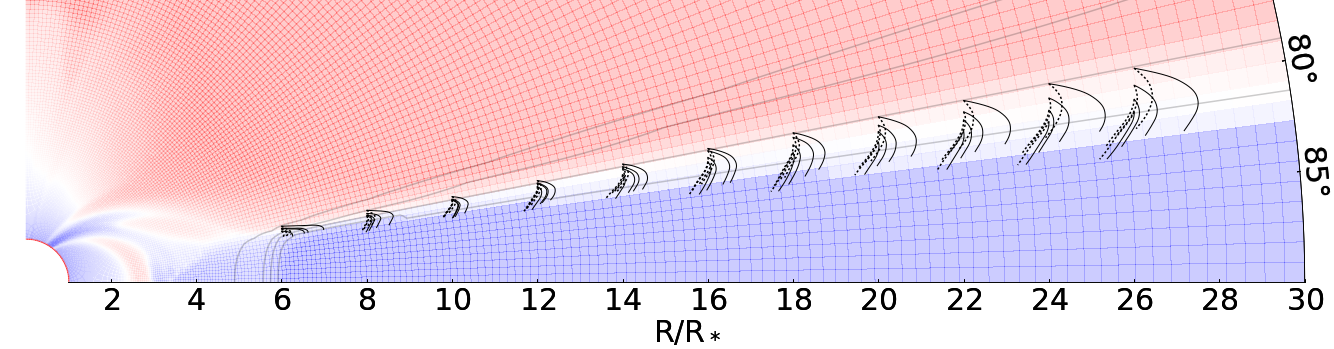}
 \includegraphics[width=\columnwidth]{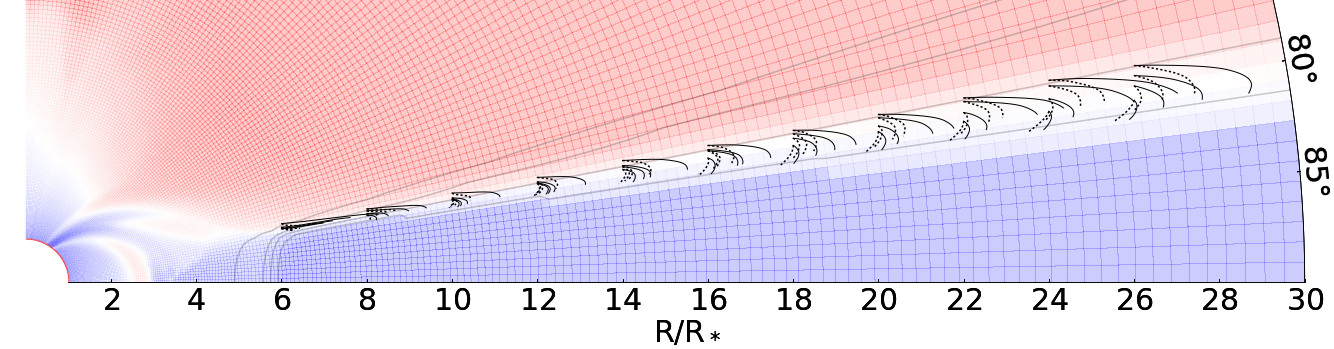}
 \includegraphics[width=\columnwidth]{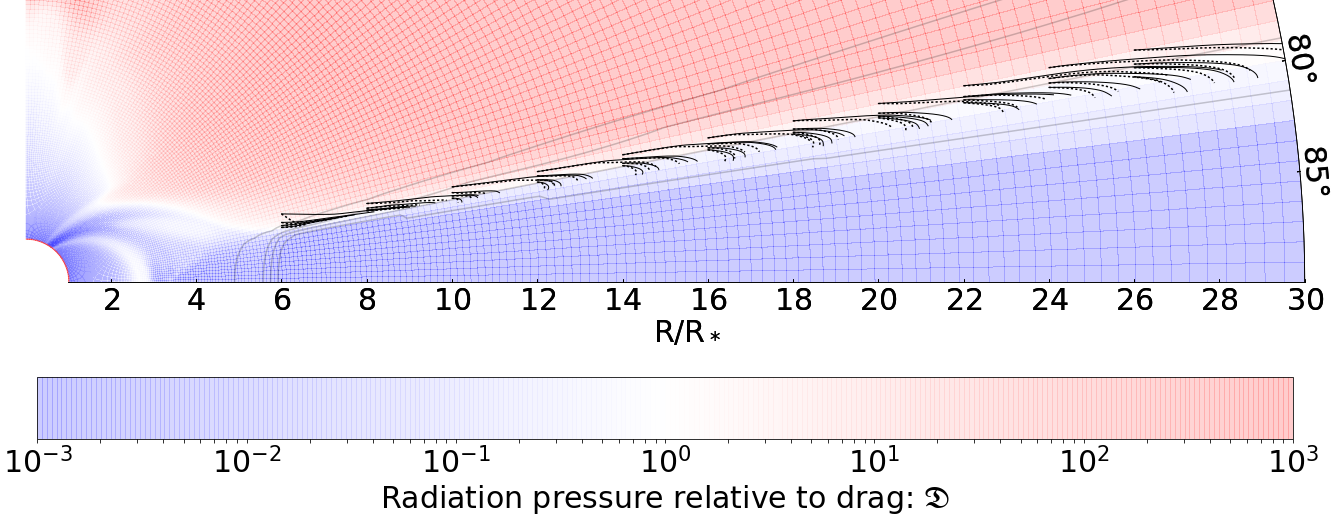}
 \caption{The same as Fig. \ref{fig:modelAsurfaceFlow}, but for the outflow model-B.  }
 \label{fig:modelBsurfaceFlow}
\end{figure}

\begin{figure}
 \includegraphics[width=\columnwidth]{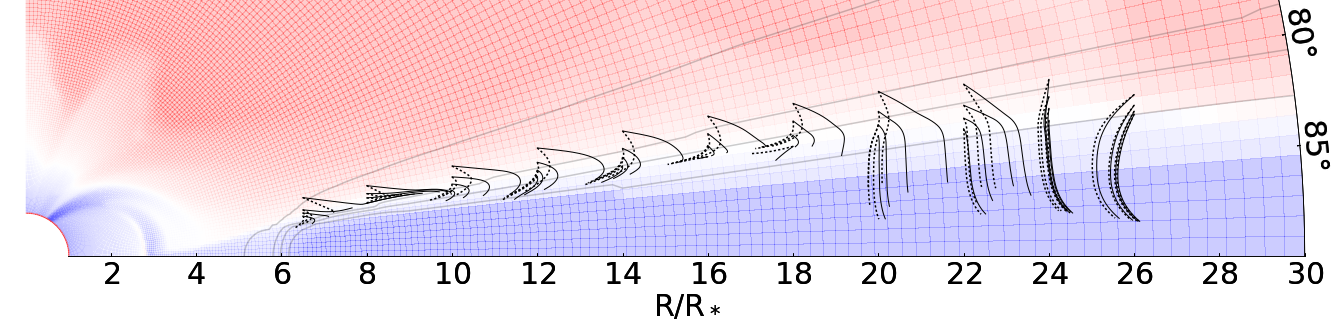}
 \includegraphics[width=\columnwidth]{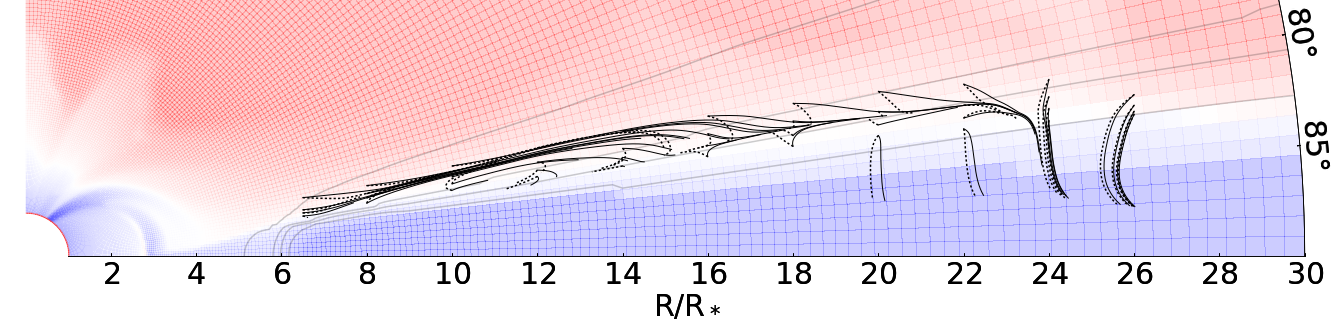}
 \includegraphics[width=\columnwidth]{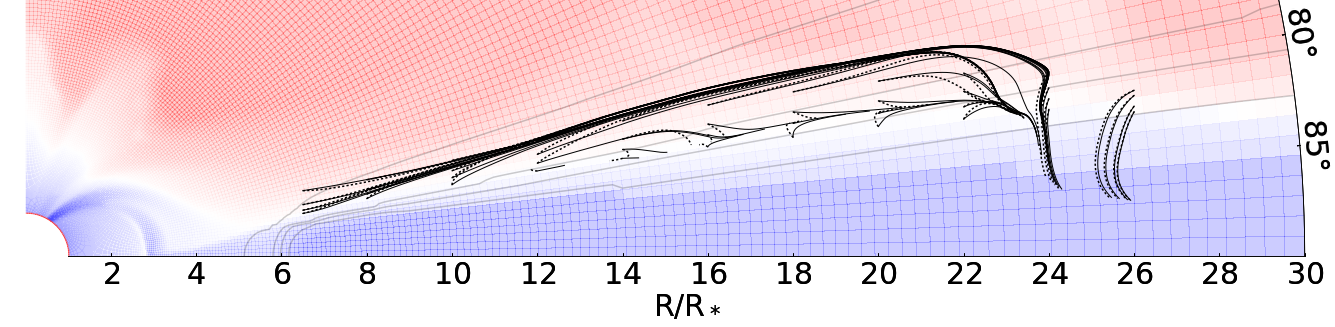}
 \includegraphics[width=\columnwidth]{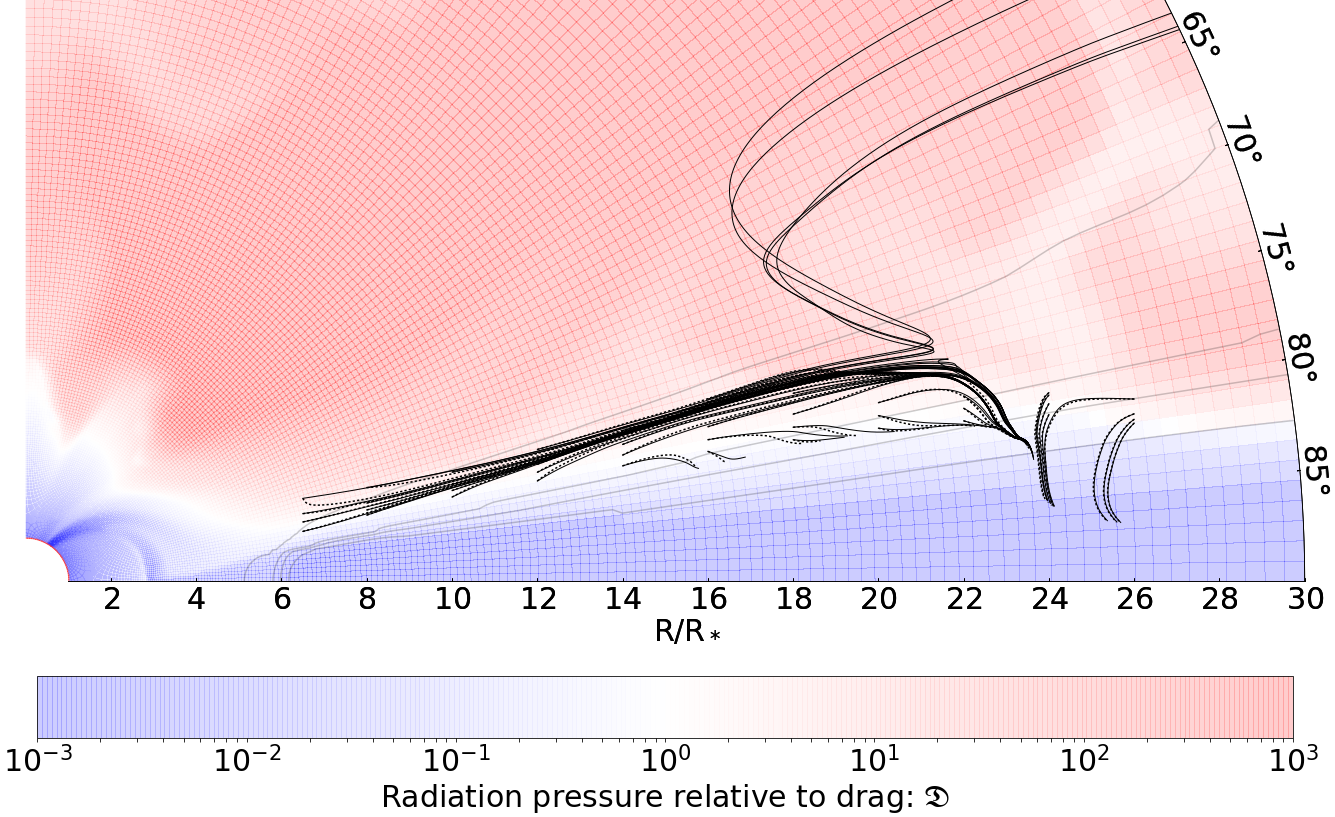}
 \caption{The same as Fig. \ref{fig:modelAsurfaceFlow}, but for the outflow model-H.  }
 \label{fig:modelHsurfaceFlow}
\end{figure}

\subsubsection{The model-B flow}

The MHD model-B has a smooth wind profile that does not create disturbance in the disc surface (Fig.\ref{fig:GasVelocity}). Hence, the dust trajectories shown in Fig.\ref{fig:modelBsurfaceFlow} are very much aligned with the predictions by \citet{Takeuchi}, combined with the gas outflow on the top of the disc surface. The big grains ($a=2\mu$m) experience only a slight extra push by the radiation pressure force compared to the gas drag by the wind alone, while the fluffy and 
sub-micron grains have a strong push by the radiation pressure. In other words, the radiation pressure force and the gas wind are skimming off dust particles from the disc surface. This means that small dust particles are quickly removed from the disc surface where the radiation pressure and/or the outflow disc wind dominate. This removal also contributes to the dust mixing within this disc region.

\subsubsection{The model-H flow}

The MHD model-H shows a very dramatic disc and wind gas dynamics, with the gas flow experiencing local direction flow reversals (Fig.\ref{fig:GasVelocity}). The dust paths show that the dust follows a myriad of trajectory shapes (see Fig.\ref{fig:modelHsurfaceFlow}) that depend on intricate details of the interplay between the radiation pressure and the gas drag driven by a complicated wind flow. The big $a=2\mu$m grains experience a significant radiation pressure drift in this model in comparison to the disc wind alone. In is interesting how the grain porosity plays a significant role in this model as the solid grains sink into the disc not too far from the starting point, while the fluffy grains glide over the disc surface far away thanks to the radiation pressure force. The modelled small grains of 0.2$\mu$m and 0.1$\mu$m radius show even more interesting behaviour. They are captured by the gas wind, which enables them to fly far away over the disc surface, until they get pulled down at about \hbox{24 $R_*$}. The radiation pressure provides an extra lift to the small grains and helps them fly even farther. In extreme situations some grains mange to reach trajectories that launch them far above the disc and then far away into the outer disc regions. Without radiation pressure such grains would stay close to the disc surface. Hence, the model-H demonstrates how a combination of radiation pressure and the MHD winds can produce large scale mixing in protoplanetary discs.

\begin{figure}
 \includegraphics[width=\columnwidth]{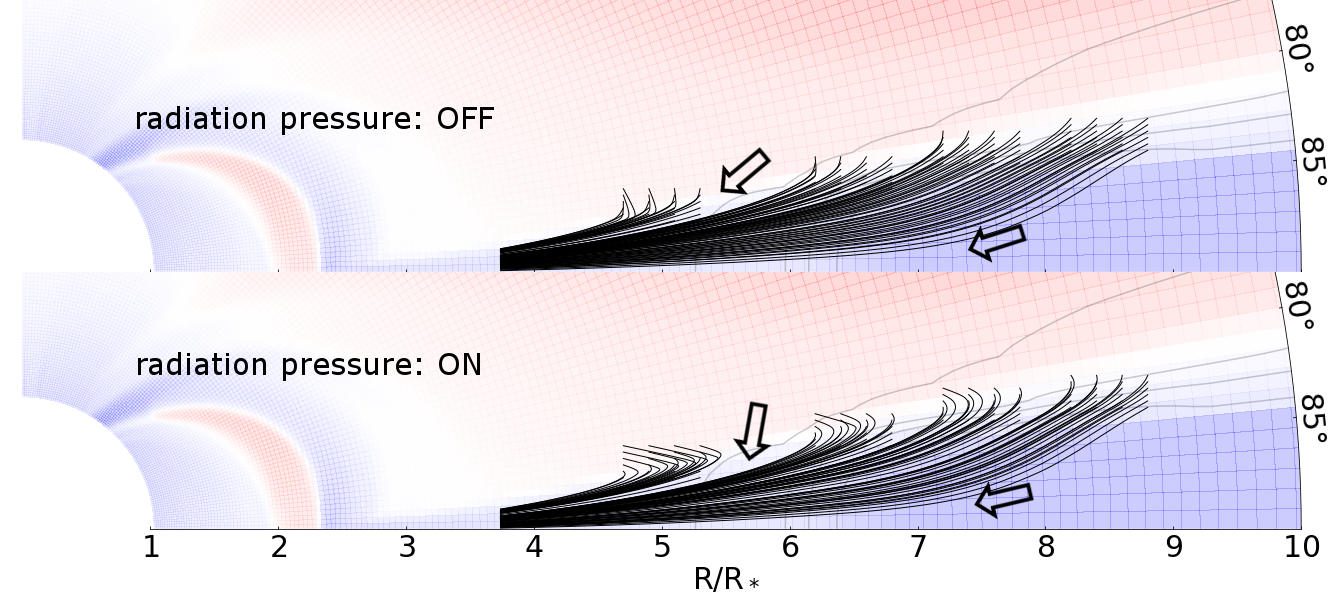}
 \includegraphics[width=\columnwidth]{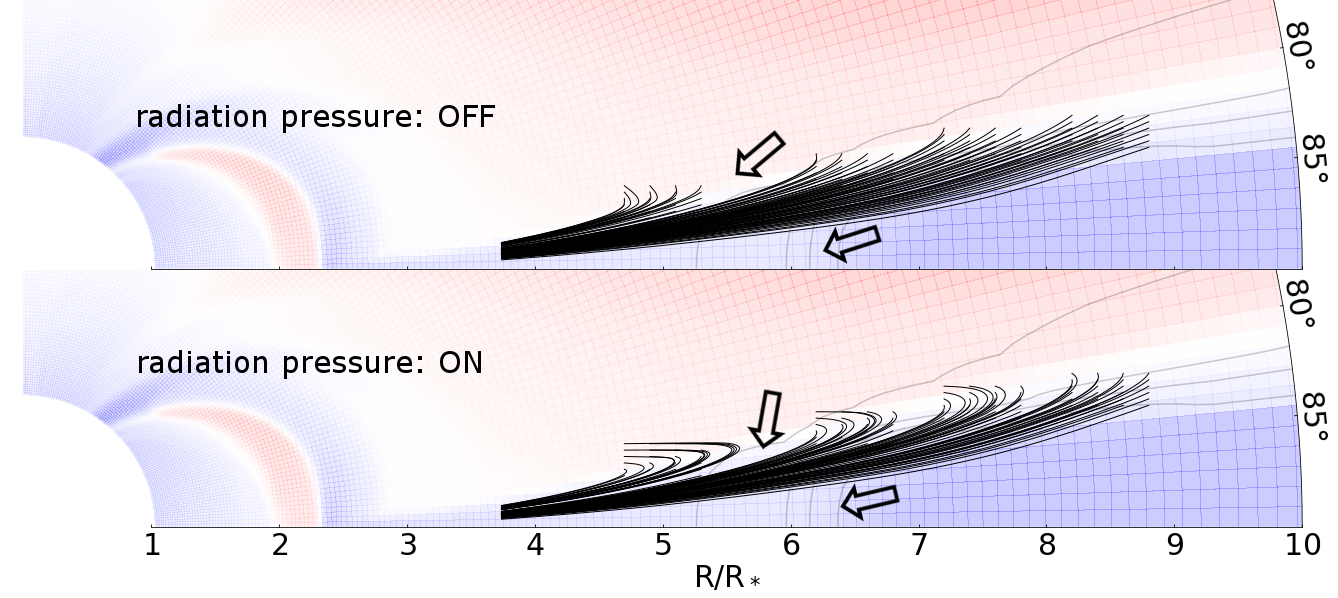}
 \includegraphics[width=\columnwidth]{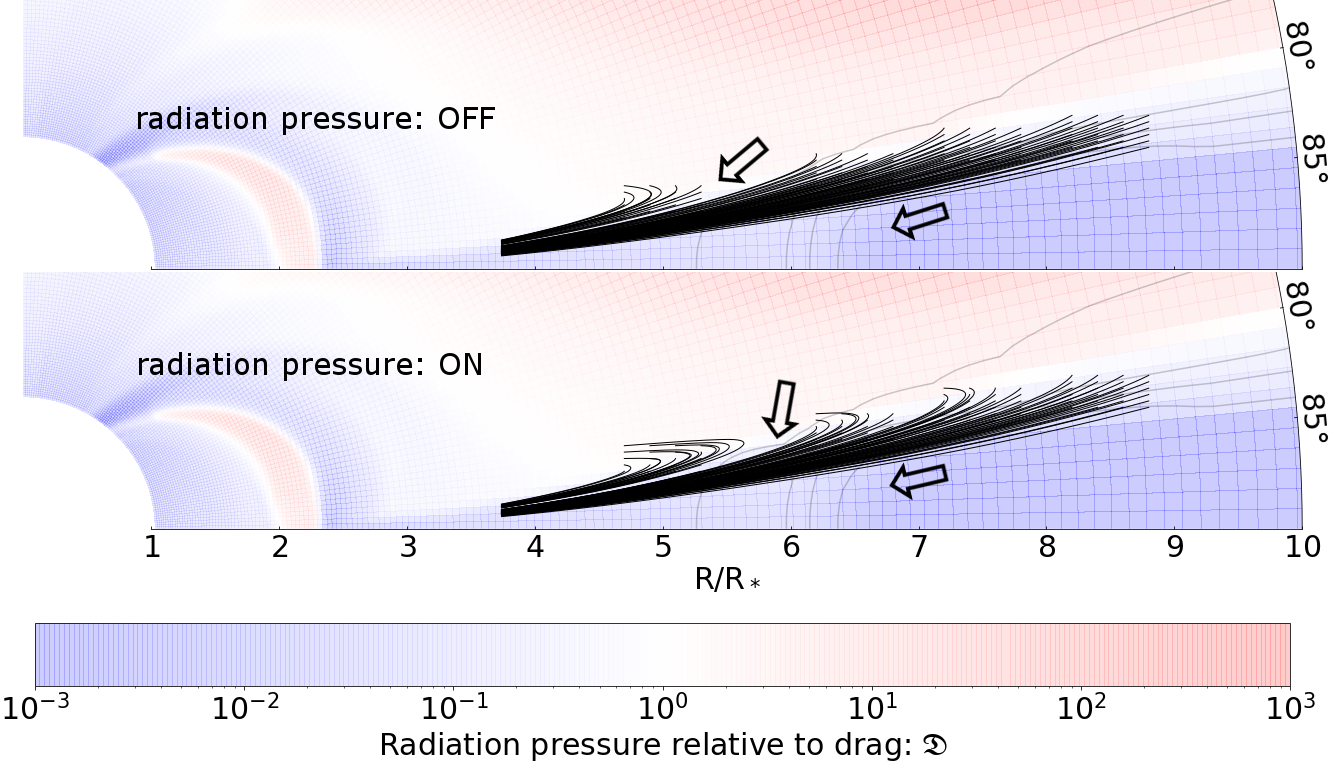}
 \caption{Trajectories of dust grains at different locations around the inner dusty disc edge in the outflow model-A. See Fig. \ref{fig:GasVelocity} for a view of the gas velocity flow. The colour map shows the strength of radiation pressure relative to the gas drag ($\mathfrak{D}$, equation \ref{eq:D}). The thin solid lines are contours of the optical depth attenuation factor $\varepsilon$ described by equation \ref{eq:f}, with values set to 0.1, 0.4, 0.7 and 0.99 (the maximum is 1.0 in regions without dust). The thick solid lines are paths of dust particles, with arrows indicating the dust flow direction for easier understanding. The radiation pressure is included or turned off for comparison, as indicated in the images. The top and bottom pair of images have the bulk density of dust grains of 3000 kg/m$^3$, while the middle pair has 1000 kg/m$^3$. The grain radius in the top two image pairs is 2$\mu$m, while the bottom pair has 0.2$\mu$m.}
 \label{fig:modelAinnerFlow}
\end{figure}

\begin{figure}
 \includegraphics[width=\columnwidth]{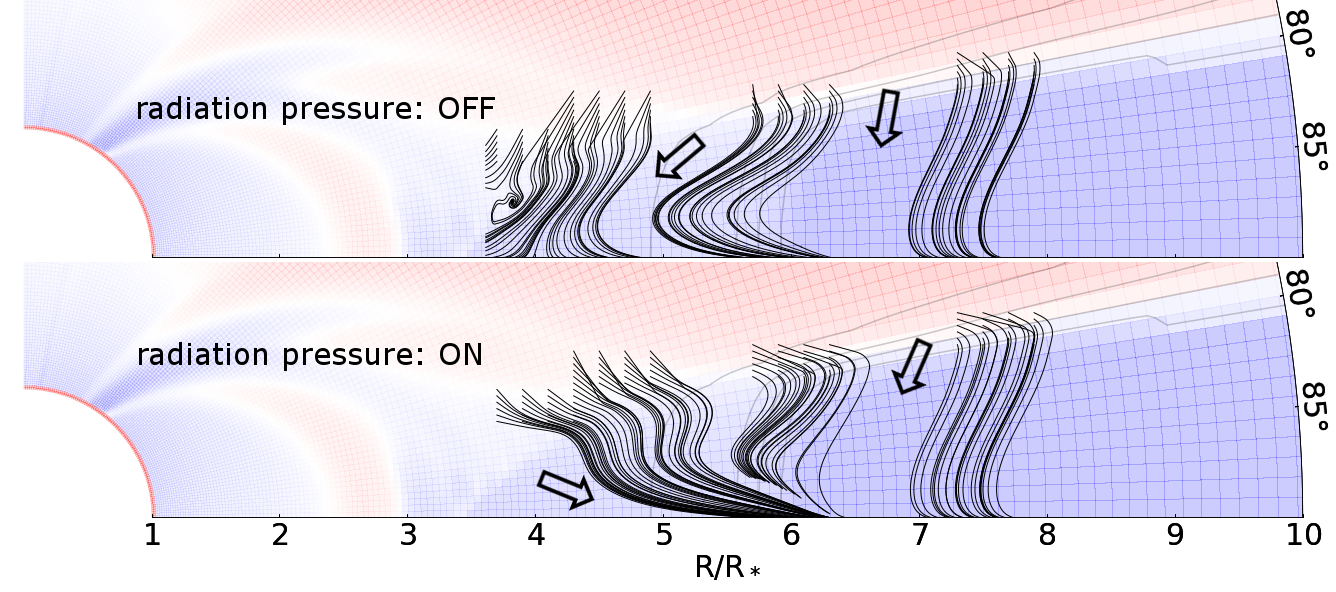}
 \includegraphics[width=\columnwidth]{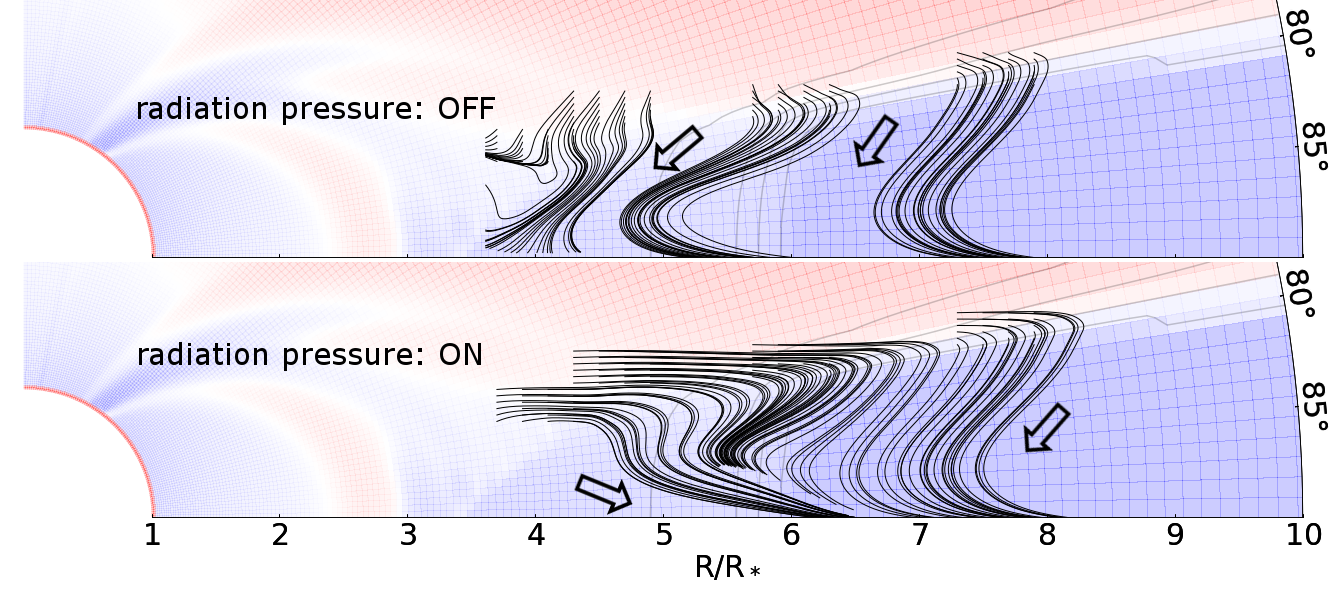}
 \includegraphics[width=\columnwidth]{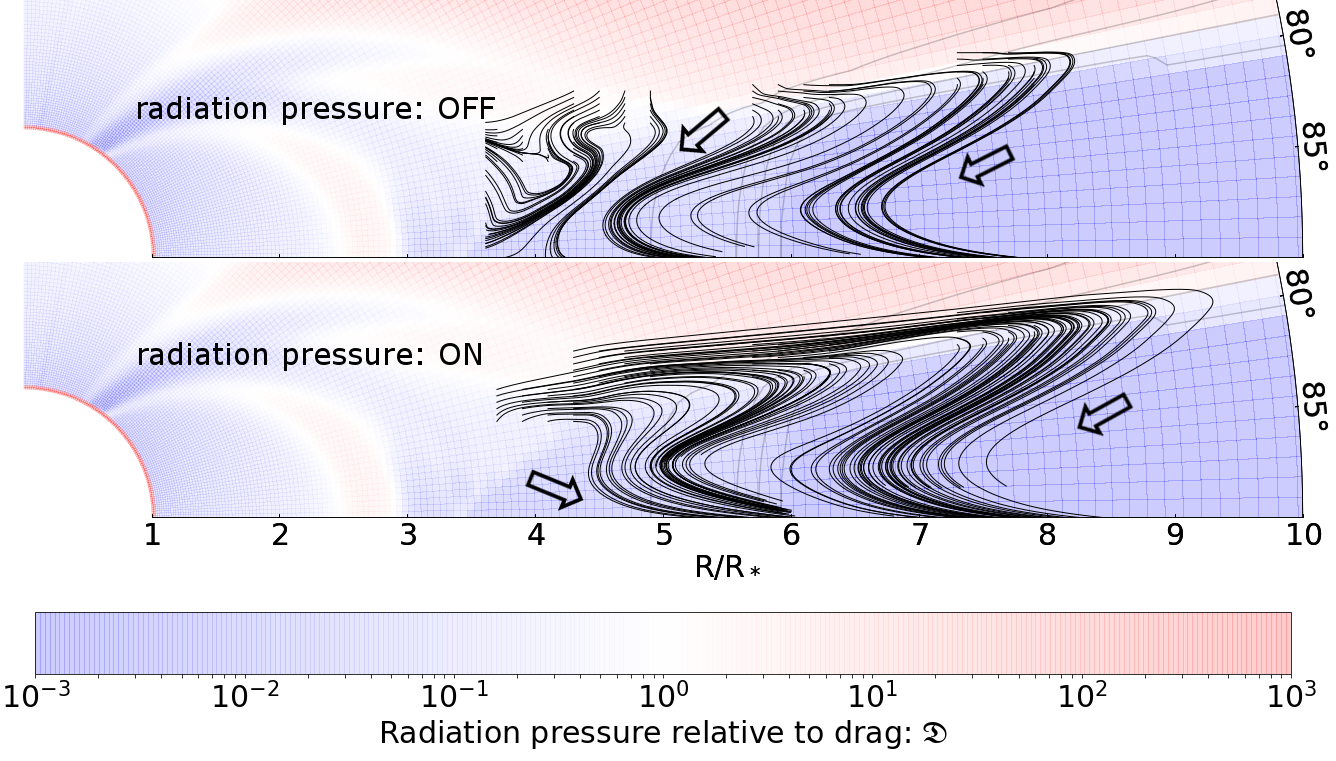}
 \caption{The same as Fig. \ref{fig:modelAinnerFlow}, but for the outflow model-B.  }
 \label{fig:modelBinnerFlow}
\end{figure}

\begin{figure}
 \includegraphics[width=\columnwidth]{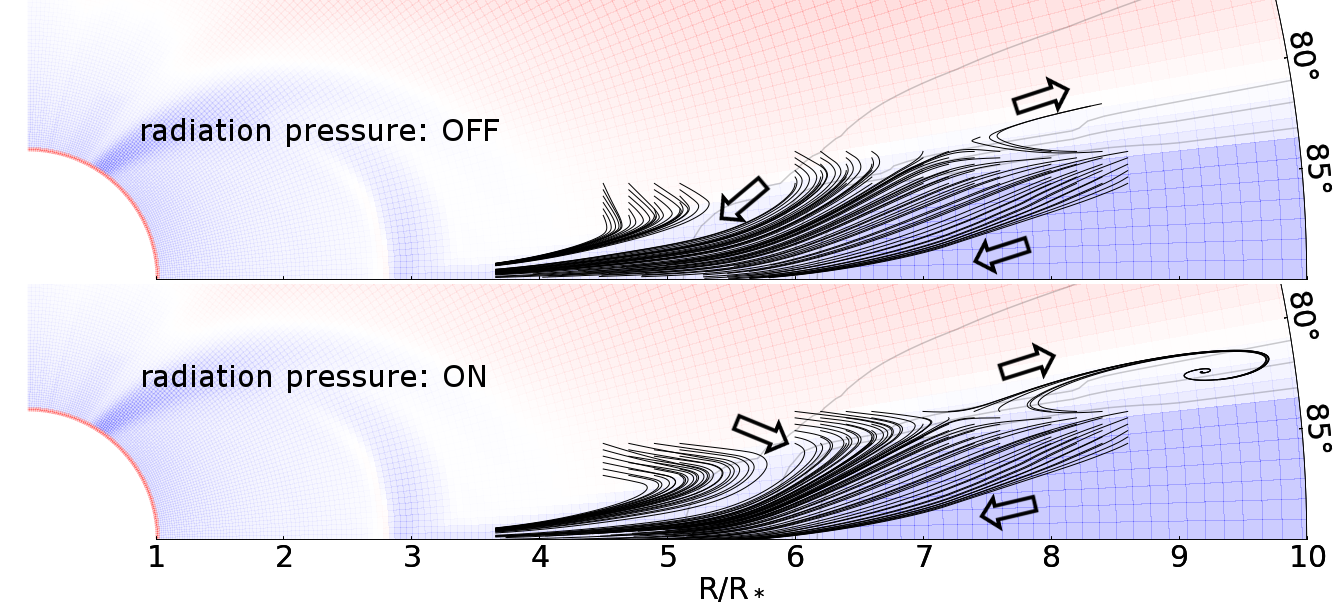}
 \includegraphics[width=\columnwidth]{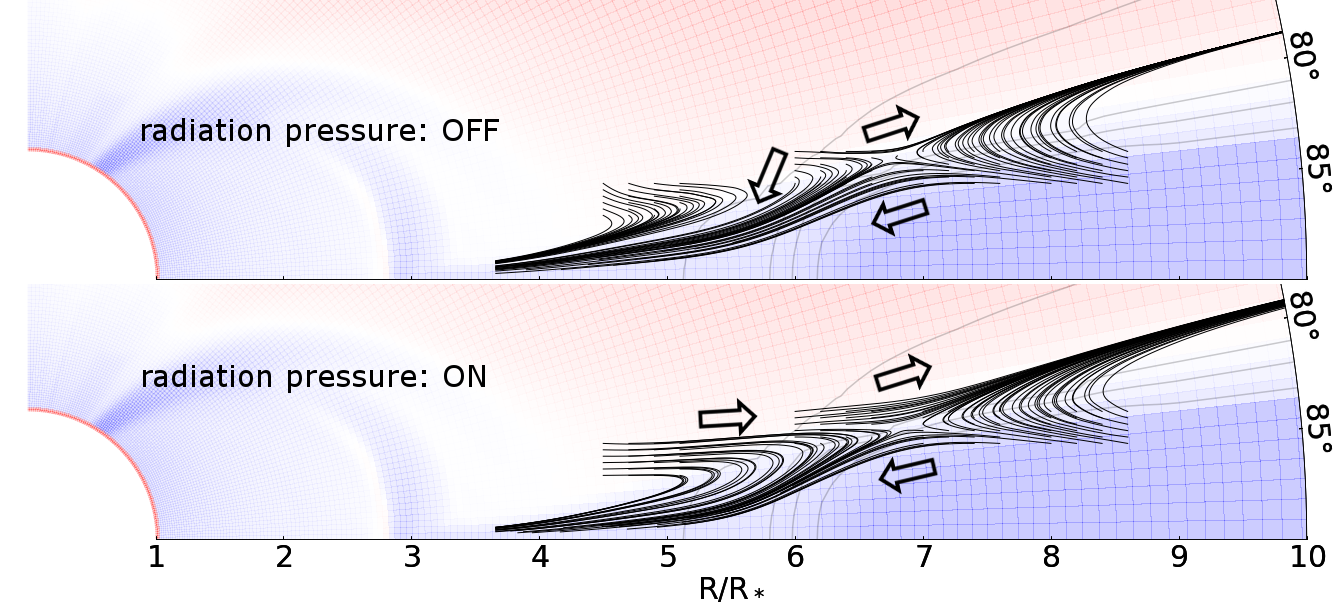}
 \includegraphics[width=\columnwidth]{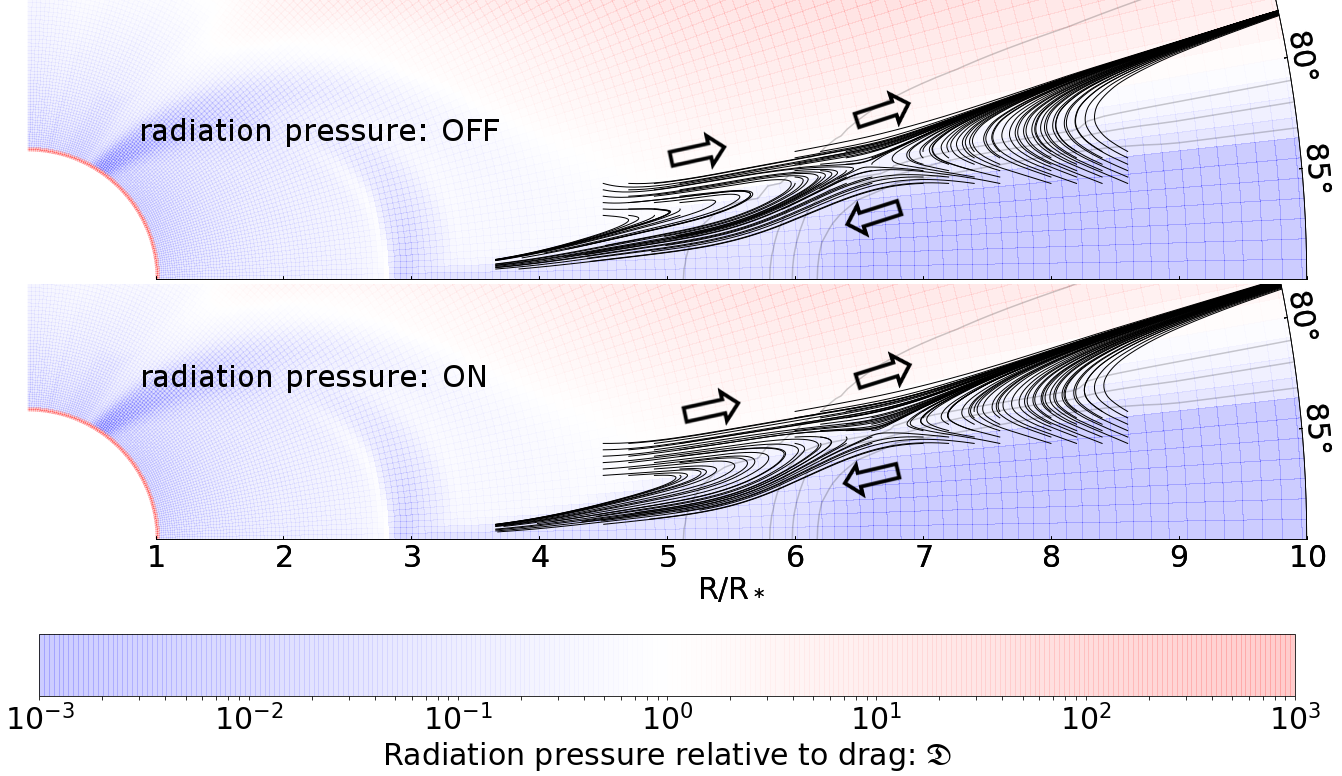}
 \caption{The same as Fig. \ref{fig:modelAinnerFlow}, but for the outflow model-H.  }
 \label{fig:modelHinnerFlow}
\end{figure}

\subsection{The disc's inner edge flow}

In our modelling we can explore the dust dynamics scenarios within the inner disc rim region (i.e. between  $R^{thick}_{in}$ and $R^{thin}_{in}$) and see how the complexity of MHD gas flow affects the dust trajectories. This region between the two (thin and thick) cut-off radii is optically thin and only grains with the efficient radiative cooling (i.e. larger grains, which is in our case sizes larger than about 0.4$\mu$m; see Fig.\ref{fig:beta}) can survive so close to the star. We position grains at various locations and track their trajectories to see where the grains will end up. The calculation is stopped if a grain exits the computational domain: reach the midplane or get closer than $R^{thin}_{in}$ to the star or go beyond 10 stellar radii. We also perform the same calculations with the radiation pressure force turned off in order to clearly see differences in the dust flow caused by the radiation pressure.

\subsubsection{The model-A inner edge flow}

Dust flow in the inner optically thin part of the model-A disc is dominated by the disc gas inflow (Fig.\ref{fig:modelAinnerFlow}). The dust trajectories display a monotonically smooth flow toward the star, with bigger grains sinking deeper into the disc. All the test particles released within the disc surface end up reaching the optically thin inner disc edge. Inclusion of the radiation pressure force into the dynamics has the effect of compressing the dust flow closer to the midplane. Smaller particles are first pushed outward if they happen to appear in the zone above the dense gas disc, but then they join the compressed flow dictated by the gas of high density. Thus, we can conclude that radiation pressure is an important factor in shaping the height of the inner disc rim. 

\subsubsection{The model-B inner edge flow}

The model-B shows a very intriguing dust flow in the inner edge region (Fig.\ref{fig:modelBinnerFlow}). The flow is very uneven, changing its directions in various ways. One reason for this is the dense gas outflow in the disc midplane (see Fig.\ref{fig:GasVelocity}). This outflow disturbs the trajectories of inward sinking dust and pushes them away from the $R^{thin}_{in}$ where the dust could sublimate. This means that all the inflowing dust accumulates in the disc midplane instead of being destroyed by the stellar heat. Moreover, the dust is pushed even further back into the optically thick disc by the radiation pressure force. 

The final outcome is that all the dust that manages to reach the optically thin zone between $R^{thick}_{in}$ and $R^{thin}_{in}$ eventually accumulates in the midplane behind the optically thick wall at $R^{thick}_{in}$. The stellar radiation is attenuated enough at that location to make the radiation pressure not important. Even though the midplane gas outflow  slowly moves that dust deeper into the disc, we see how the whole process accumulates dust grains in a vary small part of the disc midplane, which can have significant consequences for planet formation scenarios. 

The dust above the dense disc experiences outflow trajectories. In the case of smaller particles the trajectories go quite far away from the inner disc edge. The particles eventually reenter the dense disc and then start to flow inward, driven by the gas drag in the upper layers of the disc. When the inflow particles encounter the dense gas outflow in the midplane, or become exposed to the radiation pressure, their path is again switch to outflow that eventually ends up in the midplane behind the optically thick inner disc edge. Thus, in this model the radiation pressure plays two important roles: it creates dust outflow over the disc surface and it prevents dust from reaching the optically thin disc zone close to the star where it could sublimate. 

\subsubsection{The model-H inner edge flow}

The dust flow in model-H has two outcomes defined by the initial position of the released dust particles (Fig.\ref{fig:modelHinnerFlow}). The first outcome addresses dust grains deep in the disc's optically thick interior. Their trajectories resemble the solution described in model-A: a smooth flow toward the star, with the radiation pressure compressing the flow closer to the midplane. The second outcome concerns dust in the disc surface. Those grains are moving inward after the initial release, but at one point change their direction and get ejected far back above the disc surface. The consequence of such a flow is the disc surface erosion caused by both the gas wind outflow and the radiation pressure force. This dust ejection can create an optically thin halo above the disc, which has been invoked previously in order to explain the anomalously high near infrared disc emissions \citep{VinkovicHalo}.

The outflow is more pronounced in the case of fluffy grains or smaller grains. It shows how the grain porosity can play an important role in the efficacy of this outflow scenario. Notice that a collisional destruction of fluffy grains would produce small grains that are then also very efficiently pushed far back over the disc surface. Another interesting outcome is that the dusty disc would not have a sharp inner edge, but rather a funnel-like shape with the narrow end at the optically thin sublimation radius. However, different parts of the "funnel" are populated by different types of grain sizes under optically thin conditions. 

\section{Discussion}
\label{sec:discussion}

There are multiple lines of argument for dust grains existing above the dense gas and dust disc. The most convincing indication comes from the optical and infrared studies of spectral and photometric variability of young stellar objects \citep[e.g.][]{Cody,Rice}. Irregular reddening episodes are associated with variable dust extinction on orbital timescales that indicate a dynamic environment where clouds of dust move in front of the star. It would be easier to understand this phenomenon if the discs are highly inclined and clouds are actually confined to the hydrostatic disc and its turbulence, but inclinations are often such that the disc's geometrical thickness is too small to support this assumption \citep{Vinkovic07,Cody}. 

Another argument for dust above disc comes from the study of near infrared (NIR) flux. The amount of emitted NIR flux from protoplanetary discs has been a matter of controversy because it is often larger than the emission implied by the hot disc area. Since the dusty disc is truncated due to the dust sublimation, the stellar radiation can heat the dust within the entire disc's vertical profile of the disc inner rim. This expands the disc rim above the typical vertical disc size \citep{Flock17}, which then increases the emitted NIR flux. The problem is that the typical physical processes (viscous heating, turbulent diffusion, dust settling, gas pressure gradient, radiation pressure) are not capable of enlarging the dusty disc rim height to the size suitable for explaining the anomalously large fluxes \citep{Vinkovic14}. Indeed, the trajectories of dust grains in our models with radiation pressure show how the optically thick disc wall is eroded at its top. 

However, a small amount of optically thin dust above the disc would be enough the boost the NIR flux \citep{VinkovicHalo}. Possible mechanisms for creating this halo above the disc could be dust ejection by MHD disc winds or jets \citep{BansWind,Liffman20}. Our dust trajectories confirm a possibility to have dust launched above the disc and form an optically thin extension of the dusty disc rim. This might be further enhanced by stellar outburst events, when the disc surface dust is puffed away by the disc wind \citep{Abraham}. Radiation pressure by infrared photons emitted from the hot dusty rim can also help in transporting dust into such an optically thin halo \citep{Vinkovic09}.

The composition of cometary dust also favours some kind of a dust transport that would happen above the disc, with grains from the inner parts of the disc ending up in the outer parts \citep{Brownlee,Ogliore}. The meteoritic composition fits into this scneario, too. The melted or crystallised inclusions are found mixed with the more primitive, colder matrix in chondrite meteorites. This situation could be explained by disc wind transferring outward the inner disc material transformed by high temperatures \citep{Salmeron}. Isotopic studies of variability among Solar System objects also support an outward transport of thermally processed disc material \citep{Schiller}.

Both the radiation pressure force \citep{Takeuchi,Vinkovic09} and the disc wind \citep{BansWind,Giacal} have been already considered as the engine behind this transport. Results of our study demonstrate how these two mechanisms operate jointly to produce a more complicated dusty outflow. Even though dust grains can be lifted by the disc wind and survive in the optically thin space above the disc, the radiation pressure force acts on this dust very efficiently. Hence, neither the radiation pressure nor the disc wind can be used alone for modelling dust grain dynamics.

This means that the search for a good theoretical description of the protoplanetary dusty disc geometries is still open.  The work on such modelling is difficult for several reasons. The radiation pressure force is heavily dependent on the dust characterisation, from chemical to physical properties. We touched upon only two parameters: dust size and porosity. Both have significantly affected the results. 
The disc wind modelling is also a highly complex issues. In our numerical MHD solutions we rely on alpha-viscosity prescription. It is readily related to the available analytical solutions and previous work in the literature, and can benefit from the parameter study conducted in \cite{Cem19}. In the next refinement, viscosity should be related to MRI as in \cite{Flock13}, with inclusion of the disc microphysics \citep{Baietal16} and thermochemistry \citep{Wangetal19}, with radiative transfer.

In the study by \cite{Giacal} that also explored dust trajectories, the wind is produced by radially self-similar, semi-analytic computations of steady-state accretion disc threaded by a large scale magnetic field, following the work by \cite{Teitler}. Centrifugally driven wind is launched by a magnetic field, transporting the excess angular momentum outwards. Such solutions might result in unrealistic disc parameters, as e.g. low values of thermal to magnetic pressure ratio in the disc equatorial plane, or very short disc lifetimes caused by the large inflow speed. They depend on assumptions that are not needed in our global numerical simulations, initialised with a full 3D analytical solutions for a hydro-dynamical accretion disc by \cite{KK00} and self-consistently evolved in time with the addition of magnetic field.

We have not explored possibilities for more resilient dust approaching the star even further or condensing dust out of the outflow wind as it cools. Such a scenario has been proposed for launching calcium-rich, aluminum-rich inclusions (CAIs) particles by X-wind or the centrifugal interaction between the solar magnetosphere and the inner disc rim \citep{Liffman16}. This dust would enter the hottest part of the disc, truncated by the magnetospheric accretion flow. We consider this region to be too hot for any dust because of additional sources of heating that would prevent dust to survive or condense. The same objection has been raised to the CAI formation in the X-wind model \citep{Desch}. 

Another way how dust can move close to the star is when the gas density is high enough to produce significant opacity that reduces the stellar heating of the dust. The largest column gas density is along the disc mid-plane, which would reduce the mid-plane radiation pressure effects and the mid-plane dust temperature. Both effects lead to dust moving closer to the star.

Dust dynamics can be further improved by considering dust charging, where charged dust reacts to the local magnetic field. We expect dust grains to loose or gain electrons by grain-grain collisions, grain-electron interactions and by grain-photon interactions. The efficiency of each process depends on grain chemistry, shape, local gas plasma properties and photon (UV and X-ray photons, cosmic rays) abundance (affected by disc opacity). Obviously, this is a very difficult problem to solve, but if a model is build \citep[e.g.][]{Thi2019} then the Lorentz force on dust grains can be added into the dust dynamics equations. 

A more realistic description of inner disc regions also involves improvements in MHD models. For example, we keep $\alpha_v$ constant. Since the dynamic viscosity $\eta_v\propto \rho\alpha_v$ \citep[see equation A8 in][]{Cem19}, we end up with the viscosity profile defined by the density geometry of the disc. But at the same time this density would be changed just behind the inner dusty rim where the ionisation drops and therefore the local value of $\alpha_v$ would be reduced, too. In our simulations, near the disc inner rim equations are solved in the ideal MHD regime, controlled by the beta plasma value to limit the reach of diffusive terms to magnetically dominated part of the computational box. More realistic approach is taken in \cite{Flock17}, with MRI taken into account.

\section{Conclusions}
\label{sec:conclusion}

We have explored dust dynamics in the hottest parts of protoplanetary discs, where dust grains move under the influence of highly complex time-dependent non-gravitational forces. In addition to gravity and gas drag forces, we emphasised the role that the radiation pressure plays in reshaping the dust trajectories. 
The main challenge in this type of studies is devising details of the environment in which the dust trajectories are reconstructed. In our work we utilised results from two lines of research - the MHD disc wind simulations and the high-resolution multigrain 2D radiative transfer in optically thick dusty discs. We devised a computational method that combines these two sets of numerical results to set the stage for dust dynamics equations. 

We presented results from our initial exploration where we used an approximation of time-independent gas velocity fields derived from the snapshots of steady-state MHD solutions. These are then used for setting up the gas density and velocity distributions. This approach enables insights into what kind of dust trajectories are possible and how the dust dynamics reshapes the geometry of dusty discs. The impact of radiation pressure is easily visualised by turning its contribution on and off in the equations of dust movement. 

The obtained dust flight paths are often highly irregular and very sensitive to the intricate interplay between the disc, dust and wind properties. The trajectories take a simple form only in the zone of high gas density, where the gas drag dominates. Under such conditions the dust settling mimics the gas flow, but even then the radiation pressure force compresses the flow when the grains exit the optically thick part of the disc. Trajectories of grains that approach the regions of lower gas density are easily disrupted by the radiation pressure force. The final outcome of such paths is a competition between the disc wind influence by gas drag and the radiation pressure force. In some cases this ends up with grains launched high above the disc, with a trajectory that will take them far away from the star. In other cases the push is more gentle and grains do not relocate to far distances. 

However, it is always the case that the radiation pressure force skims dust from the surface of dusty discs. The disc wind helps this process and often guides grains into more complex trajectories, which precludes making a generic model of dusty outflow density structure. The picture is further complicated when different types of grains are considered. We showed how smaller grains and grain porosity can enhance the dust outflow. All these scenarios qualitatively agree with the complex spectral and photometric variability of such young stellar objects. 

Further work is needed to build more comprehensive models based on our methodology. Our next step is to introduce time-dependent disc wind fields and observe how grains behave when dragged around by a more realistic quasi-steady state MHD wind behaviour. We will also address the impact of radiation pressure on dusty discs in Herbig Ae stars and how the infrared radiation from hot dust contributes to the radiation pressure force \citep{Vinkovic09}. In this paper we showed how radiation pressure is an important contributor to dust dynamics in T Tau stars, but in more luminous pre-main-sequence stars its importance grows significantly as it scales with the square root of luminosity. For example, in Herbig Ae stars the radiation pressure force can be about 3 to 10 times stronger than in T Tau stars, which can dramatically reshape the inner dusty disc structure around Herbig Ae stars \citep{Vinkovic14}.

\section*{Acknowledgements}

Authors acknowledge collaboration with the Croatian project STARDUST through HRZZ grant IP-2014-09-8656. M\v{C} developed the setup for star-disc simulations while in CEA, Saclay, under the ANR Toupies grant. His work in NCAC Warsaw is funded by the Polish NCN grant No. 2019/33/B/ST9/01564, and in Shanghai Observatory by CAS-TWAS President's Fellowship for
International PhD Students grant No. 2020VMC0002. We thank ASIAA/TIARA (PL and XL clusters) in Taipei, Taiwan and NCAC (PSK and CHUCK clusters) in Warsaw, Poland for access to Linux computer clusters used for the high-performance computations, and the {\sc pluto} team for the possibility to use the code. DV acknowledges Technology Innovation Centre Me\dj{}imurje for logistical support.

\section*{Data availability}

The simulated data in this paper were created using the pseudo-code described in Section \ref{subsec:algorithm}. Our Python implementation of this pseudo-code will be shared on reasonable request to the corresponding author. The algorithm requires import of simulation results from MHD simulations. For this we used already published results from previous research as described in Section \ref{Sec:MHD}.








\bsp	
\label{lastpage}
\end{document}